\documentclass[10pt]{article} % use larger type; default would be 10pt

\usepackage[utf8]{inputenc} % set input encoding (not needed with XeLaTeX)

\usepackage{geometry} % to change the page dimensions
\usepackage{graphicx} % support the \includegraphics command and options

\usepackage{booktabs} % for much better looking tables
\usepackage{array} % for better arrays (eg matrices) in maths
\usepackage{paralist} % very flexible & customisable lists (eg. enumerate/itemize, etc.)
\usepackage{verbatim} % adds environment for commenting out blocks of text & for better verbatim
\usepackage{subfig} % make it possible to include more than one captioned figure/table in a single float
\usepackage{fancyhdr} % This should be set AFTER setting up the page geometry
\usepackage{sectsty}
\allsectionsfont{\sffamily\mdseries\upshape} % (See the fntguide.pdf for font help)
\usepackage[nottoc,notlof,notlot]{tocbibind} % Put the bibliography in the ToC
\usepackage[titles,subfigure]{tocloft} % Alter the style of the Table of Contents

\begin{document}

\title{Shadows of magnetically charged rotating black holes surrounded by quintessence}

\author{Chengxiang Sun$^{1}$, Yunqi Liu$^{1,2,3}$, Wei-Liang Qian$^{2,4,1}$,Ruihong Yue$^{1,3}$ \thanks{yunqiliu@yzu.edu.cn(Corresponding author),~wlqian@usp.br,~rhyue@yzu.edu.cn}}
%\date{} % Activate to display a given date or no date (if empty),
         % otherwise the current date is printed 

 \maketitle

\begin{center}
\textit{1. Center for Gravitation and Cosmology, College of Physical Science and Technology, Yangzhou University, Yangzhou 225009, China}\\
\textit{2. Escola de Engenharia de Lorena, Universidade de S\~ao Paulo, 12602-810, Lorena, SP, Brazil}\\
\textit{3. School of Aeronautics and Astronautics, Shanghai Jiao Tong University, Shanghai 200240, China}\\
\textit{4. Faculdade de Engenharia de Guaratinguet\'a, Universidade Estadual Paulista, 12516-410, Guaratinguet\'a, SP, Brazil}
\end{center}

\begin{center}
\date{Nov. 5th, 2021}
\end{center}

\begin{abstract}
In this work, we study the optical properties of a class of magnetically charged rotating black hole spacetimes. 
The black holes in question are assumed to be immersed in the quintessence field, and subsequently, the resulting black hole shadows are expected to be modified by the presence of the dark energy.
We investigate the photon region and the black hole shadow, and in particular, their dependence on the relevant physical conditions such as the state parameter of the quintessence, the angular momentum, and the magnitude of the magnetic charge.
It is shown that the photon regions sensitively depend on the horizon structure and possess intricate features.
Moreover, from the viewpoint of a static observer, we explore a few observables, especially those associated with the distortion of the observed black hole shadows.
\end{abstract}

\newpage
\section{Introduction}

By employing the {\it very long baseline interferometry} technique, the image of M87* was recently captured by the Event Horizon Telescope Collaboration~\cite{Akiyama 2019}.
Besides the achievement of another milestone in observing the spacetime geometry of the strong field region in terms of the electromagnetic signals, it leads to a promising future of astrophysics.
Based on the Van Cittert–Zernike theorem, the measurements of the Fourier transform of the brightness distribution were accomplished through the correlated signals between distant observatories.
As was first pointed out in the literature by Bardeen {\it et al.}~\cite{Bardeen1973, Young1976}, certain optical features of the image, such as the black hole shadow, are entirely governed by the intrinsic black hole parameters, such as the mass, charge, and angular momentum.
As a result, they constitute a direct probe of the compact object.
On the practical side, however, at the present stage, the obtained image still largely depends on the specified structure of the accretion disk of the black hole, as well as its dynamics.
It is expected, as the {\it planetary} scale efforts continue, one might eventually acquire images with substantially better resolution, from which the photon rings can be unambiguously identified.
Accordingly, on the theoretical side, the studies of black hole shadow have triggered a wave of renewed interest~\cite{Bardeen1973,Young1976,Vires2000,Hioki2009,Abdujabbarov,Hou2018,Stuchlik2018,Jusufi,Konoplya2019,Wei2019,Roy2019,Vagnozzi2020,Chang2020,Badia2020,He2020} in recent years. 

The accelerating expansion of our Universe~\cite{Perlmutter1999, Riess1998, Garnavich1998} implies that a significant portion of the energy density is {\it dark}, namely, in the form of an exotic fluid of negative pressure. 
Apart from the cosmological constant $\Lambda$~\cite{Krauss1995, Weinberg1989}, other prescriptions of dark energy consist of different dynamic models. 
Among others, quintessence~\cite{Ratra1988,Wetterich1988,Chiba1997,Ferreira1997,Copeland1998,Caldwell:1997ii,Zlatev1999,Jassal2005} provides a dynamical description of dark energy in terms of a minimally coupled sacalr field.
Regarding the equation of state $p=\omega \epsilon$, the model's state parameter $\omega$ usually falls within the range of $-1 \le \omega \le -\frac13$.  
In particular, the case $\omega=-1$ is reminiscent of that of a constant cosmological constant. 

In this regard, the black hole metrics immersed in the quintessence and their respective properties have since been a pertinent topic of study.
In~\cite{Kiselev:2002dx}, Kiselev presented the first Schwarzschild black hole solution with the presence of the quintessence field. 
Subsequently, by employing the Newman-Janis algorithm~\cite{newmanj1965,Drake:1998gf}, Toshmatov {\it et al.} further generalized the solution to the rotating case~\cite{Toshmatov:2015npp}. 
The corresponding Kerr-Newman-AdS solution was also obtained~\cite{Xu:2016jod} by Xu {\it et al.} by adopting a similar approach. 
The above solution was refined and further generalized by Wang {\it et al.}~\cite{Wang:2019kzk} by scrutinizing the consistency regarding the black hole thermodynamics.
More recently, in~\cite{Benavides-Gallego:2018odl}, Benavides-Gallego {\it et al.} have considered a magnetically charged rotating black hole surrounded y the quintessence.
The authors further analyzed the properties of the event horizons, the ergosphere, and the angular motion of a zero angular momentum observer.

In reality, most black holes are not static but rotating and possibly charged.
In order to obtain the metric of the rotating black holes, the Newman-Janis algorithm has been extensively explored~\cite{Toshmatov:2015npp, Xu:2016jod, Benavides-Gallego:2018odl}.
The algorithm is a complexification technique for deriving the rotating counterpart from a static solution of the Einstein field equations.
The procedure mostly consists of the decomposition of the contravariant components of the static metric in the Eddington-Finkelstein coordinates in terms of a null tetrad, and a complex ``tilde'' transformation, which generates a new metric.
The essential part of the algorithm is the feasibility of encountering a proper null tetrad~\cite{Benavides-Gallego:2018odl}, while the tilde transformation might bring a certain arbitrariness~\cite{Drake:1998gf}.
Besides, the magnetically charged black holes have also been a subject of much attention.
The static metric was derived in~\cite{Nam:2018uvc}, from which interesting features have been observed at a short distance. 
The present paper involves an attempt to study the optical properties of the magnetically charged rotating black holes recently derived in~\cite{Benavides-Gallego:2018odl}.
The black holes in question are assumed to be immersed in the quintessence field, and subsequently, the resulting black hole shadows are expected to be modified by the presence of the dark energy.
It is noted that in the aforementioned studies, the authors have focused largely on the specific equation of state where $\omega=-\frac23$.
However, it has been pointed out by Abdujabbarov {\it et al.}~\cite{Abdujabbarov:2015pqp} that the shadow of the black hole might be significantly modified due to the presence of quintessence.
Therefore, among others, we are interested in exploring the dependence of the optical properties on the state parameter of dark energy in our study.

The remainder of the paper is organized as follows. 
In Sec.~\ref{models}, we briefly discuss the main characteristics of the black hole metric in question. 
We then proceed to analyze the horizon structure and the photon region as a function of the metric parameters in Sec.~\ref{P-R}. 
In Sec.~\ref{shadow} the black hole shadows are calculated and among others, their dependence on the equation of state of the quintessence field are explored. 
Moreover, in Sec.~\ref{emission}, we extend the study to possible observables, and in particular, those associated with the distortion parameters of the black hole shadow.
Further discussions and implications of the present study are given in the last section. 
\section{Black holes in Quintessential fields} \label{models}

The black hole solution concerned in this work was first introduced in~\cite{Benavides-Gallego:2018odl}.
The metric in the Boyer-Lindquist coordinates reads
\begin{eqnarray}\label{metric}
ds^2=-\frac{\Delta_r}{
\Sigma}(dt-a\sin^2{\theta}d{\phi})^2+\frac{\Sigma}{\Delta_r}dr^2+\Sigma d{\theta}^2+\frac{\sin^2{\theta}}{\Sigma}(adt-(r^2+a^2)d{\phi})^2 ,
\end{eqnarray}
where some auxiliary functions are defined as follows 
\begin{eqnarray}\label{deltar}
\Sigma&=&r^2+a^2\cos^2{\theta},\nonumber\\
\Delta_r&=&r^2+a^2-\frac{2Mr^4}{Q^3+r^3}-{\gamma}r^{-3\omega+1} ,
\end{eqnarray}
where the parameters $a$, $M$, and $Q$ denote the black hole spin, mass, and charge, respectively. 
$\Lambda$ is the cosmological constant while $\omega$ is the state parameter of the quintessence, $\gamma$ represents the intensity of the quintessence field related to the black hole. 

\begin{figure}[htb]
\centering
\begin{minipage}[c]{1\linewidth}
\centering
\includegraphics[width=5cm]{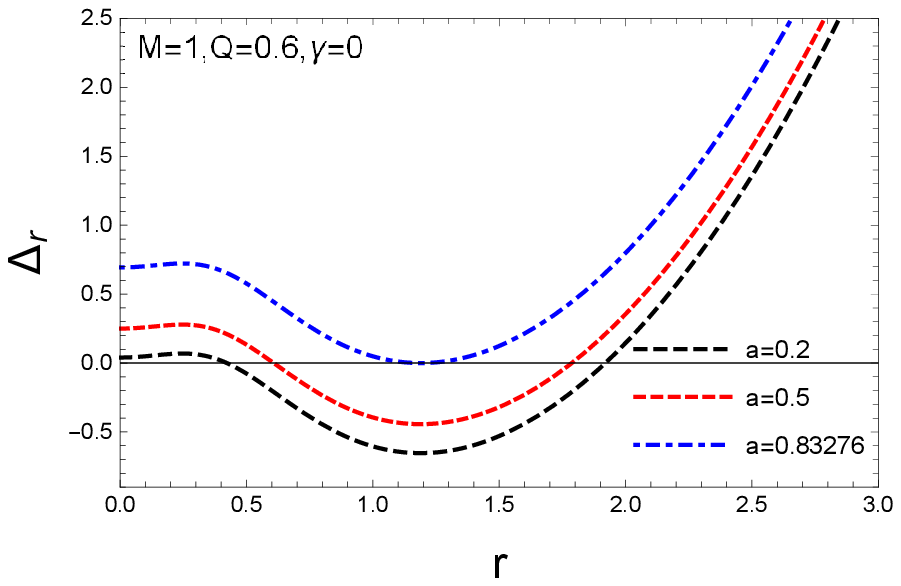}
\includegraphics[width=5cm]{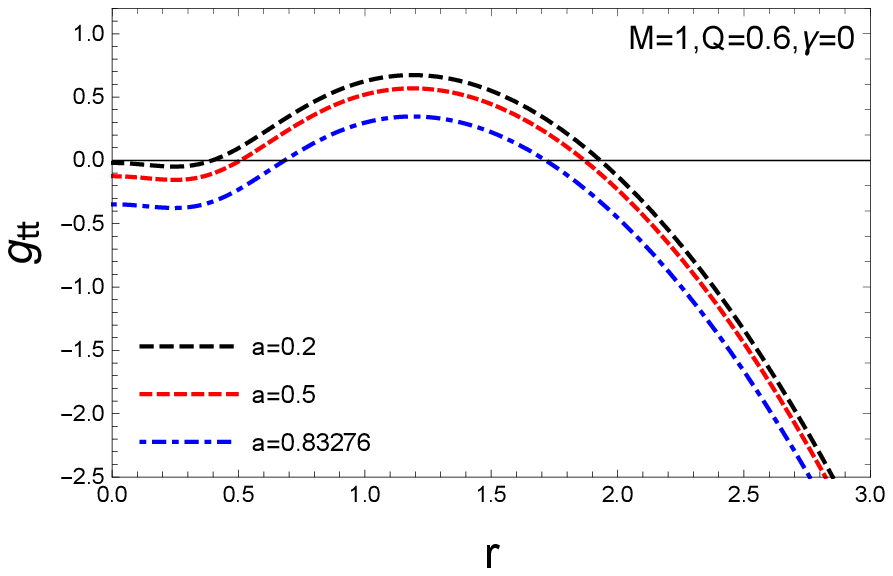}
\includegraphics[width=5cm]{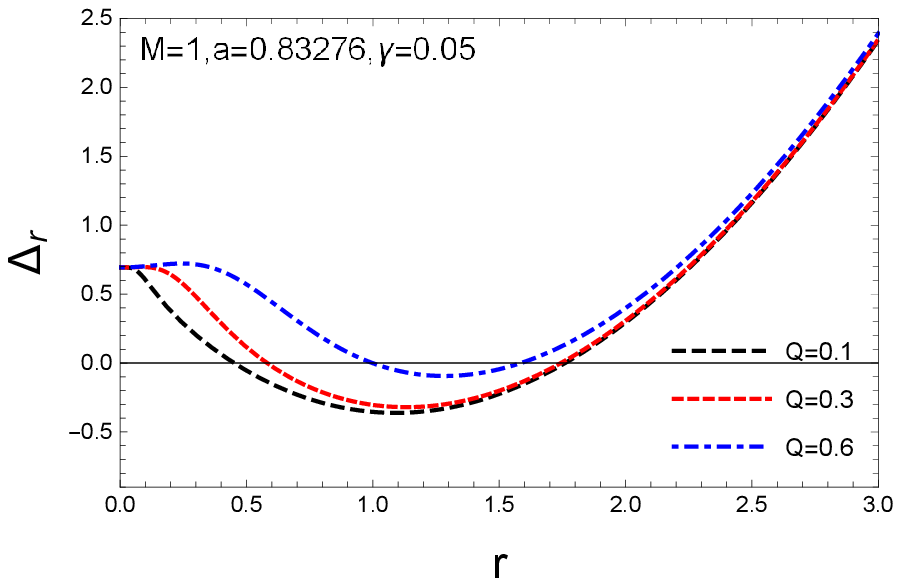}
\includegraphics[width=5cm]{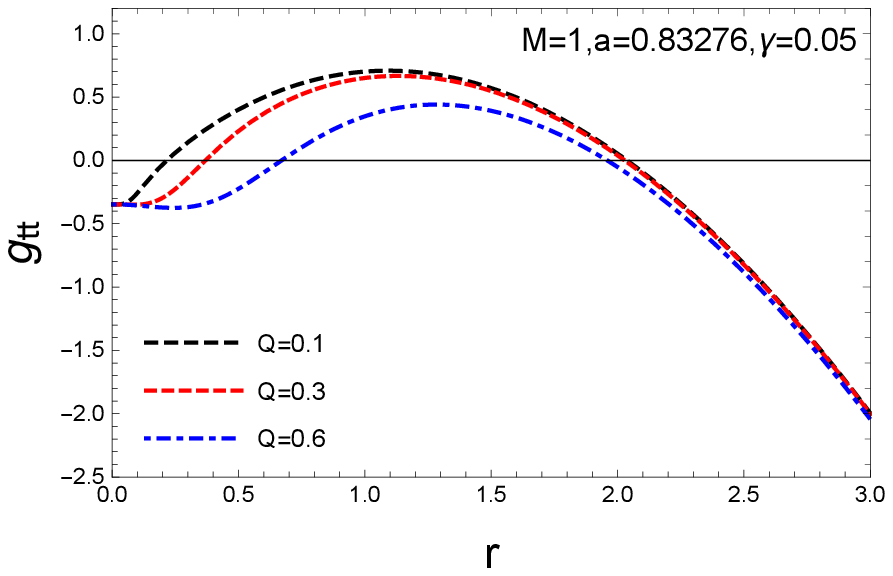}
\end{minipage}%
\caption{(Color online)  
The calculated $\Delta_r$ and $g_{tt}$ of the metric Eq.~(\ref{metric}) as functions of the radius coordinate $r$, which in turn determine the horizon and ergoregion.
For the upper row, parameters M,~Q,~$\gamma$ are fixed, the rotating parameter $a$ takes the values $0.2,~0.5,~0.83276$,
For the lower row, the charge $Q$ takes $0.1,~0.3~,0.6$, while holding the other parameters constant.
Left panel: The calculated curves of $\Delta_r$ whose roots govern the horizon structure. 
Right panel: The calculated $g_{tt}$ as functions of $r$ for $\theta=\pi/4$, which determine the corresponding ergoregion. }
\label{m2-rad}
\end{figure}

The metric describes a rotating black hole with a magnetic charge, governed by $a$, $M$, and $Q$.
It can be derived~\cite{Benavides-Gallego:2018odl} by applying a modified version of the Newman-Janis method~\cite{newmanj1965,Drake:1998gf} to a static magnetically charged black hole immersed in the quintessence field~\cite{Nam:2018uvc}.
The horizons of the black hole are determined by the roots $\Delta_r=0$.
In Fig.\ref{m2-rad}, one shows the behavior of $\Delta_r$ and $g_{tt}$ as a function of $r$. 
Besides its dependence on the angular momentum of the black hole, it can be shown~\cite{Benavides-Gallego:2018odl} that the existence of an outer horizon is rather sensitive to the value of the magnetic charge.
The shape of the ergosphere, on the other hand, depends on all the black hole parameters, except when it is located on the equatorial plane.
In the following section, we proceed to discuss the photon region of the black hole metric.

\section{Photon Region}\label{P-R}

For a black hole, the existence of the photon region is crucial for the construction of its shadow.
By definition, the region is filled up by bounded null geodesics~\cite{Grenzebach:2014fha}.
For the specific metric given by Eq.~(\ref{metric}), the feasibility of separation of variable implies that such geodesics are spherical\footnote{This is because in such cases, one can derive a one-dimensional effective potential for the radial motion, and as a result, an unstable geodesic corresponds to the maximum of the potential.}.
Subsequently, the boundary of black holes shadow is mostly determined by the unstable photon orbits around the black hole, except for the {\it metastable} ones~\cite{Qian:2021qow}. 
Indeed, when perturbed, the photons traveling along the unstable orbits of the black holes would either fall onto the event horizon or escape to infinity. 

To show the geometry near the horizon and the unstable orbits, in this section, we study the trajectories of the spherical null geodesics and the black hole photon region. 
One starts by writing down the Hamilton Jacobi equation $S$~\cite{Chandrasekhar:1984siy} which governs null geodesic in the spacetime Eq.~(\ref{metric}),
\begin{eqnarray}\label{hje-quation}
\frac{{\partial}S}{{\partial}\tau}=-\frac{1}{2}g^{\mu\nu}\frac{{\partial}S}{{\partial}x_{\mu}}\frac{{\partial}S}{{\partial}x_{\nu}} ,
\end{eqnarray}
where $\tau$ is the affine parameter, $x^{\mu}$ represents the four-dimensional coordinates $(t,r,\theta,\phi)$. 
To proceed, we assume that the solution to the Hamilton-Jacobi equation takes the form
\begin{eqnarray}\label{hj-ansatz}
S=\frac{1}{2}m^2\tau-Et+L\phi+S_r(r)+S_{\theta}(\theta) ,
\end{eqnarray}
where for time-like geodesics $m^2=1$ and for null geodesics $m^2=0$. 
The functions $S_r(r)$ and $S_\theta(\theta)$ univariable, and $E$ and $L$ are constants related to the associated time and azimuthal Killing vectors $\zeta_t$ and $\zeta_{\phi}$. 
Plugging the metric tensor of Eq.(\ref{metric}) and the ansatz Eq.(\ref{hj-ansatz}) into the Hamilton Jacobi equation Eq.(\ref{hje-quation}), one finds
\begin{eqnarray}
m^2&=&\frac{(r^2+a^2)^2-\Delta_r a^2\sin^2{(\theta)}}{\Delta_r\Sigma}E^2-\frac{\Delta_r\csc^2{(\theta)}-a^2}{\Delta_r\Sigma}L^2\nonumber\\
&&-\frac{2a(r^2+a^2)-2\Delta_ra}{\Delta_r\Sigma}EL-\frac{\Delta_r}{\Sigma}(\frac{{\partial}S_r(r)}{{\partial}r})^2-\frac{1}{\Sigma}(\frac{{\partial}S_{\theta}(\theta)}{{\partial}\theta})^2.
\end{eqnarray}
As a result, the method of separation of variables gives
\begin{eqnarray}
\frac{dS_r(r)}{dr}&=&\frac{\sqrt{R(r)}}{\Delta_r},\\
\frac{dS_{\theta}(\theta)}{dr}&=&\sqrt{{\Theta}(\theta)},
\end{eqnarray}
where
\begin{eqnarray}
R(r)&=&(aL-(a^2+r^2)E)^2-(m^2r^2+K)\Delta_r,\label{equ8}\qquad\\
{\Theta}(\theta)&=&(K-a^2m^2\cos^2(\theta))-(L\csc(\theta)-aE\sin(\theta))^2 ,\label{equ9}
\end{eqnarray}
where one introduces the Carter constant $K$.

The resulting form of $S$ reads
\begin{eqnarray}
S=\frac{1}{2}m^2\tau-Et+L\phi+\int\frac{\sqrt{R(r)}}{\Delta_r}dr+\int\sqrt{\Theta(\theta)}d\theta ,
\end{eqnarray}
which readily gives rise to the equation of motion for the geodesic as follows
\begin{eqnarray}
{\Sigma}\dot{t}&=&\frac{((r^2+a^2)E-aL)(r^2+a^2)}{\Delta_r}-a(aE\sin^2(\theta)-L),\\
{\Sigma}\dot{r}&=&\sqrt{R(r)},\label{equ10}\\
{\Sigma}\dot{\theta}&=&\sqrt{{\Theta}(\theta)},\\
{\Sigma}\dot{\phi}&=&\frac{a((r^2+a^2)E-aL)}{\Delta_r}-(aE-L\csc^2(\theta)) ,\label{equ11}
\end{eqnarray}
where a dot denotes the derivative with respect to the affine parameter $\tau$. 

It is not difficult to confirm that the bounded null geodesics consist of spherical orbits that are confined on the $r=constant$ spheres. 
The latter satisfies $\frac{dr}{d\tau}=0$ and $\frac{d^2r}{d{\tau}^2}=0$. 
Subsequently, the region filled by such null geodesics furnishes the photon region~\cite{Grenzebach:2014fha}. 
From Eq.~(\ref{equ10}), we have
\begin{eqnarray}\label{equ1}
R(r)=0,\quad\frac{dR(r)}{dr}=0 
\end{eqnarray}
with the additional condition that
\begin{eqnarray}\label{equ2}
\Theta(\theta)\geq0{\quad}as\quad\theta\in[0,\pi].
\end{eqnarray}
By solving Eq.~(\ref{equ1}), the spherical null geodesics with radius $r=r_p$ is governed by the ratios of the conserved quantities $\xi=\frac{L}{E}$ and $\eta=\frac{K}{E^2}$.
In particular, one finds
\begin{eqnarray}
{\xi}&=&\frac{r^2_p}{a}+a-\frac{4r{\Delta_r}(r)}{a{\Delta_r}'(r)},\label{equ3}\\
\eta&=&\frac{16r^2{\Delta_r}(r)}{\Delta_r'(r)^2}.\label{equ4}
\end{eqnarray}
Inserting Eqs.~(\ref{equ3}) and (\ref{equ4}) into conditi on (\ref{equ2}), one derives an inequality which determine the boundary of the photon region, namely, 
\begin{eqnarray}\label{equ5}
16r^2a^2{\Delta_r(r)}\sin^2(\theta)\geq(4r\Delta_r(r)-\Sigma{\Delta_r'(r)})^2. 
\end{eqnarray}

The stability of the spherical null geodesics with radius $r_p$ is dictated by the second-order derivative of the effective potential.
To be specific, they are unstable against the radial perturbation when $\frac{d^2R(r)}{dr^2}{\mid_{r=r_p}}>0$, stable when $\frac{d^2R(r)}{dr^2}{\mid_{r=r_p}}<0$. 
The former is relevant to the boundary of the black hole shadow. 
By using Eqs.~(\ref{equ8}), (\ref{equ3}) and (\ref{equ4}), finally one obtains
\begin{eqnarray}\label{unstable}
\frac{d^2R(r)}{dr^2}{\Big{|}_{r=r_p}}=8E^2\left(r^2+\frac{2r\Delta_r(r)}{\Delta_r'(r)}-\frac{2r^2\Delta_r(r)\Delta_r''(r)}{\Delta_r'(r)^2}\right)>0.
\end{eqnarray}

\begin{figure}[htb]
	\centering
	\begin{minipage}[c]{1\linewidth}
		\centering
		\includegraphics[width=3.3cm,height=3.3cm]{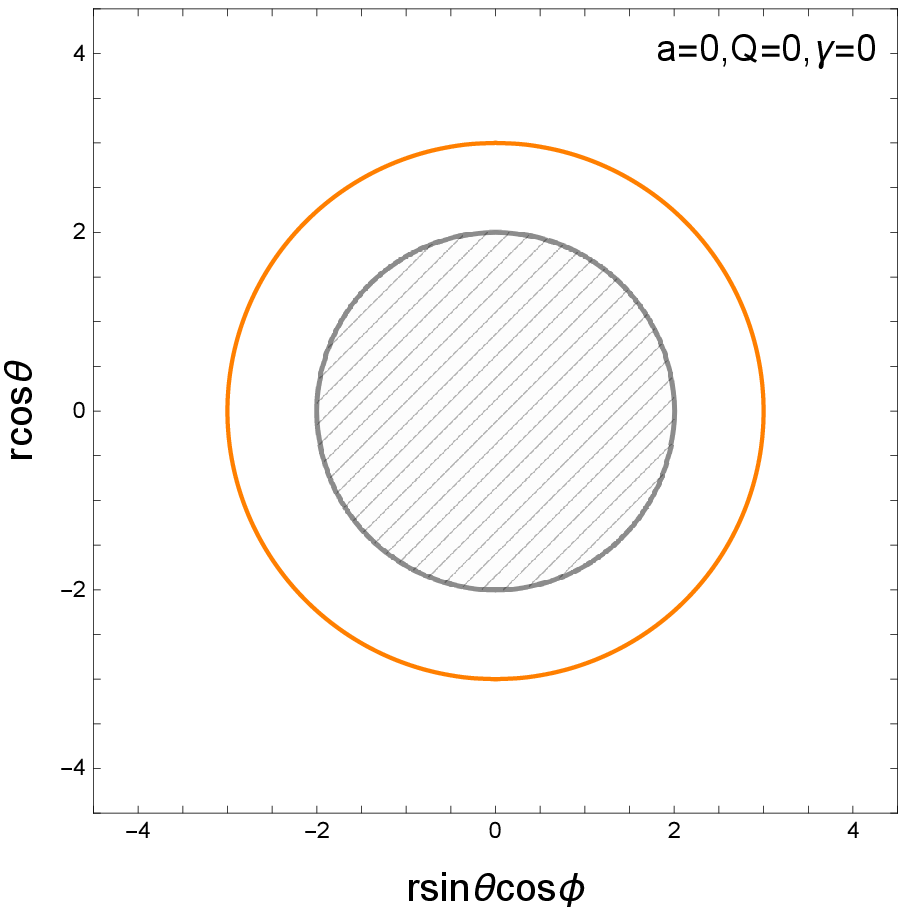}
		\hfill%
		\includegraphics[width=3.3cm,height=3.3cm]{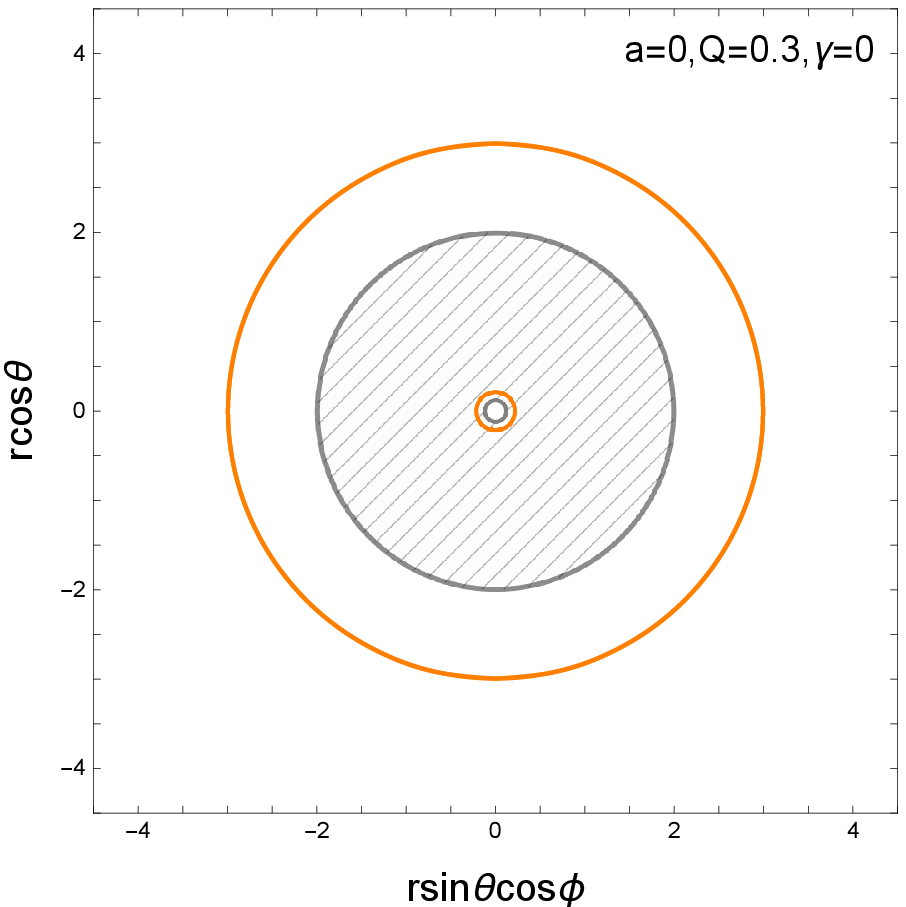}
		\hfill%
		\includegraphics[width=3.3cm,height=3.3cm]{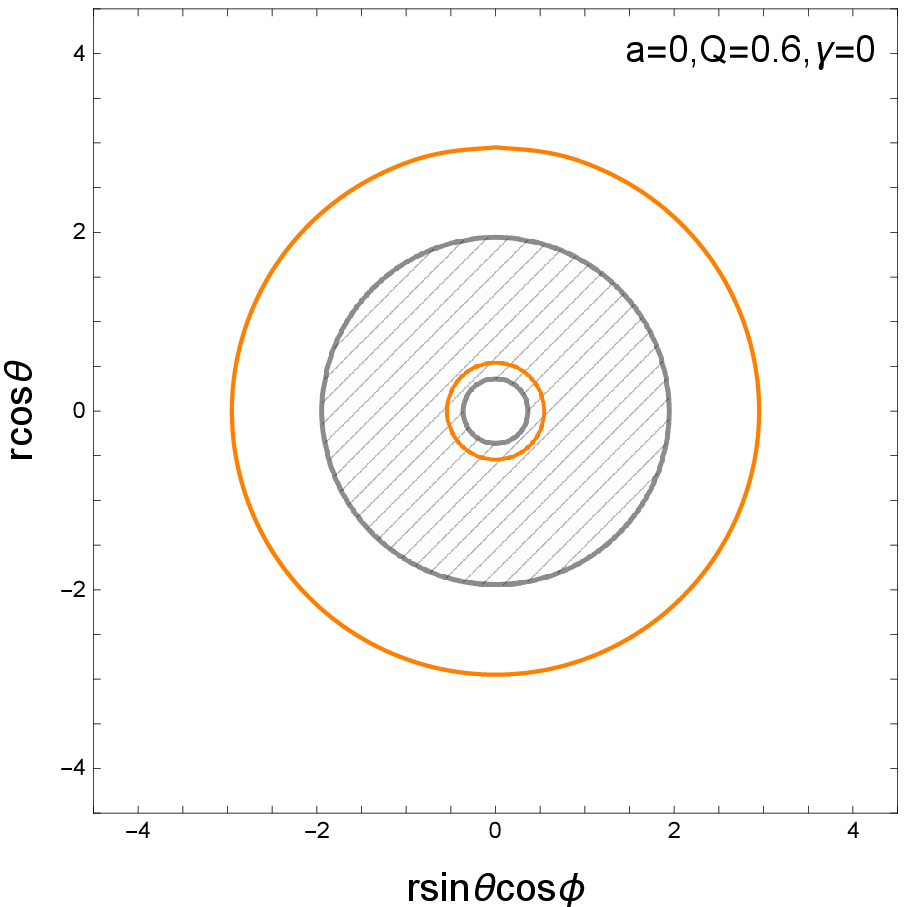}
		\hfill%
		\includegraphics[width=3.3cm,height=3.3cm]{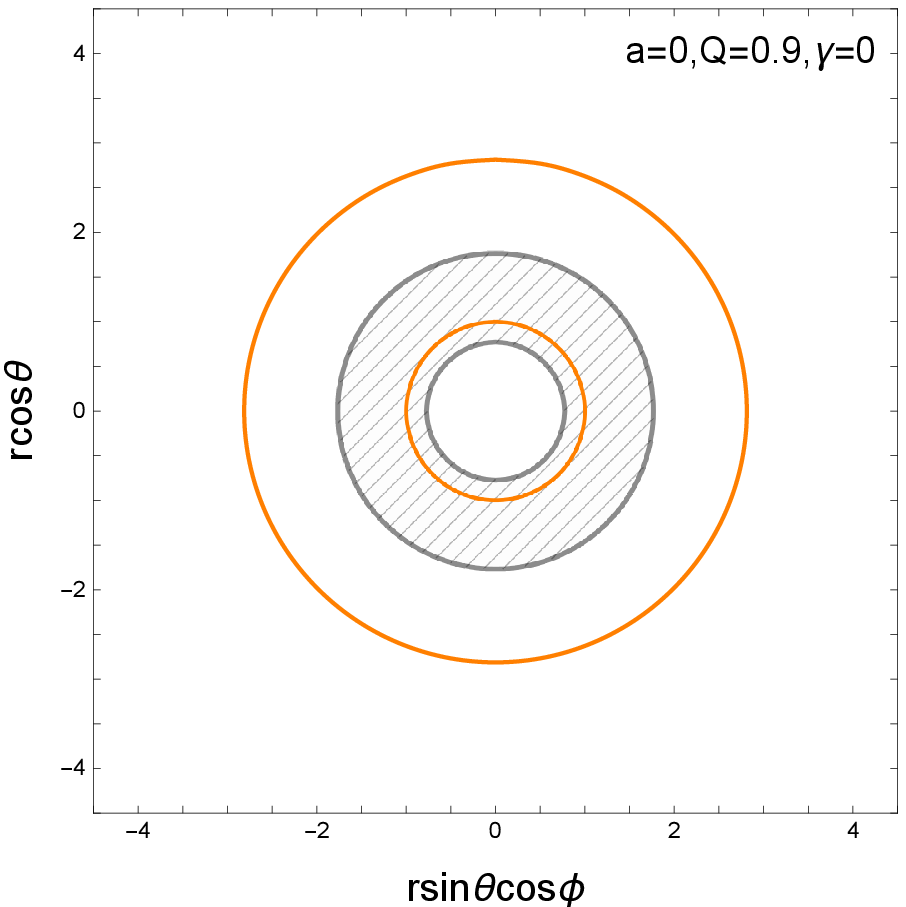}
		\vfill%
		\includegraphics[width=3.3cm,height=3.3cm]{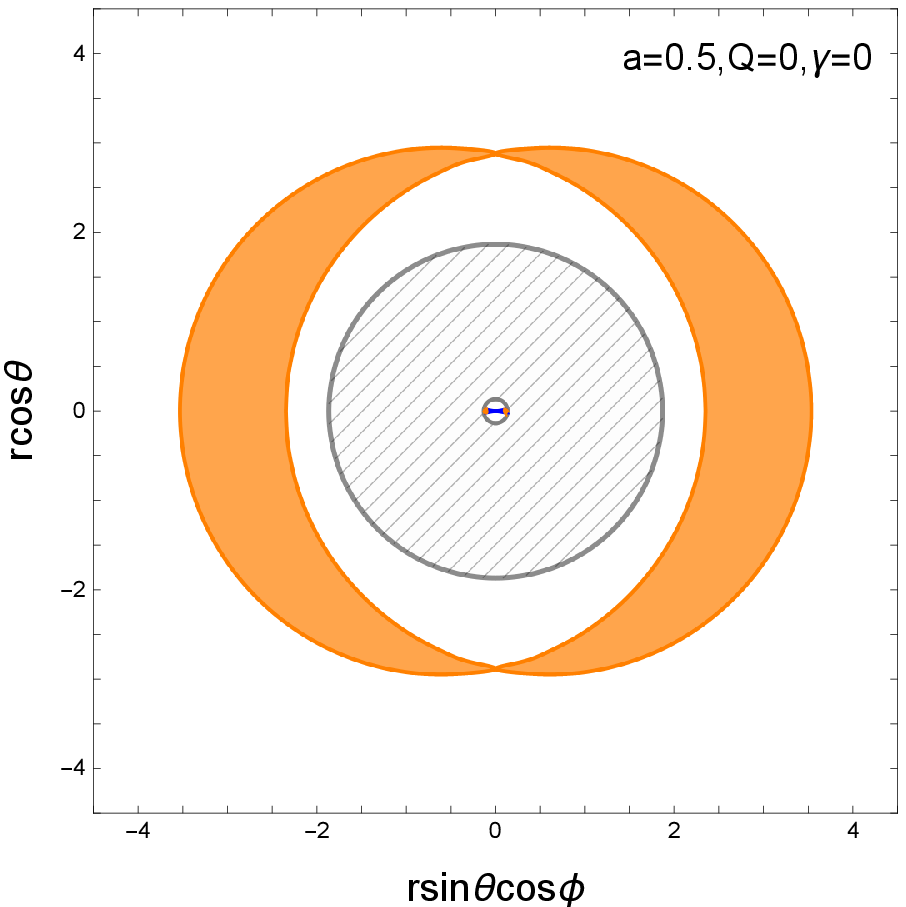}
		\hfill%
		\includegraphics[width=3.3cm,height=3.3cm]{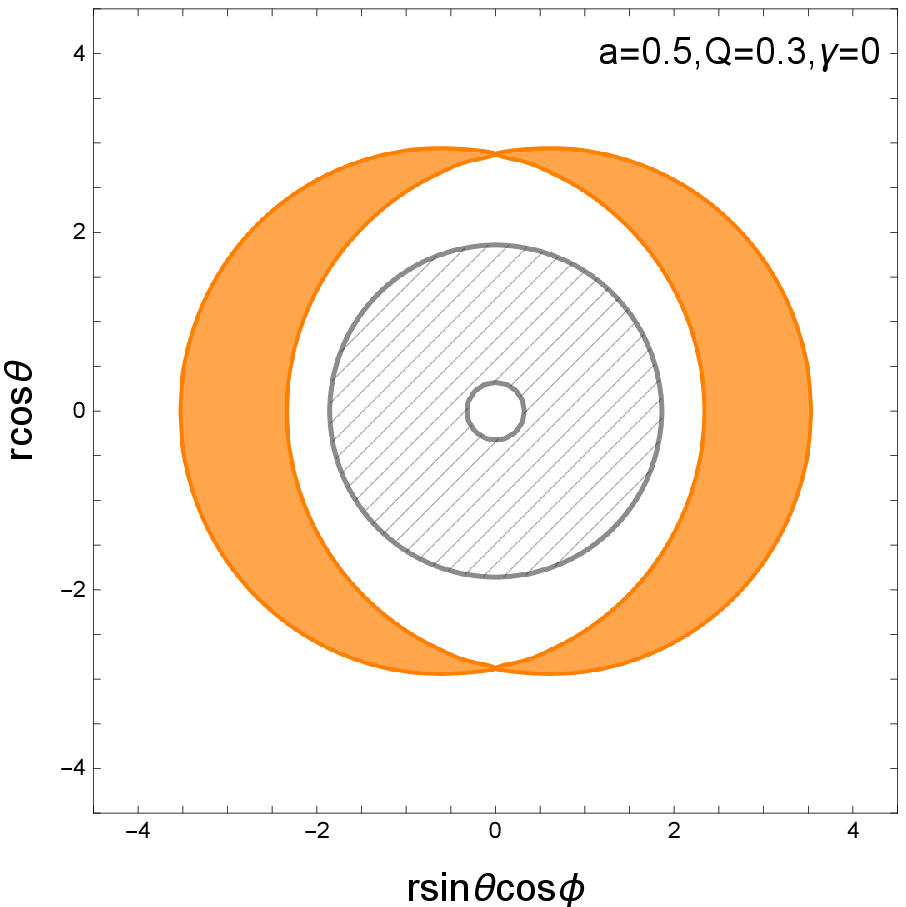}
		\hfill%
		\includegraphics[width=3.3cm,height=3.3cm]{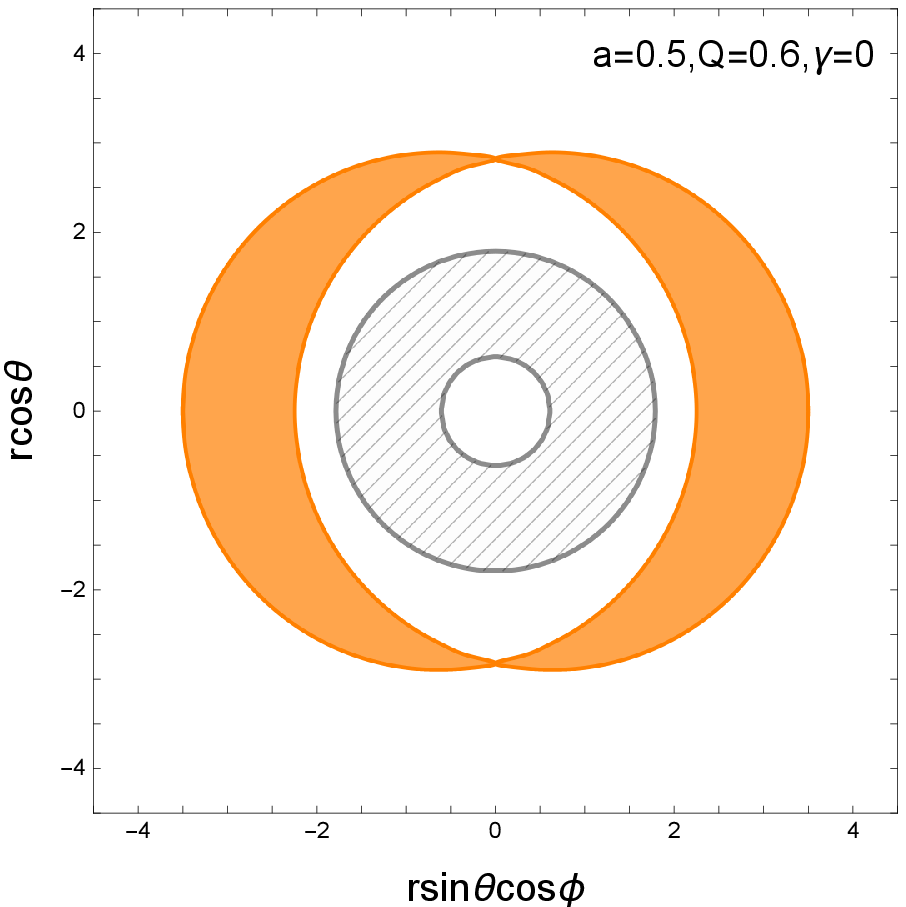}
		\hfill%
		\includegraphics[width=3.3cm,height=3.3cm]{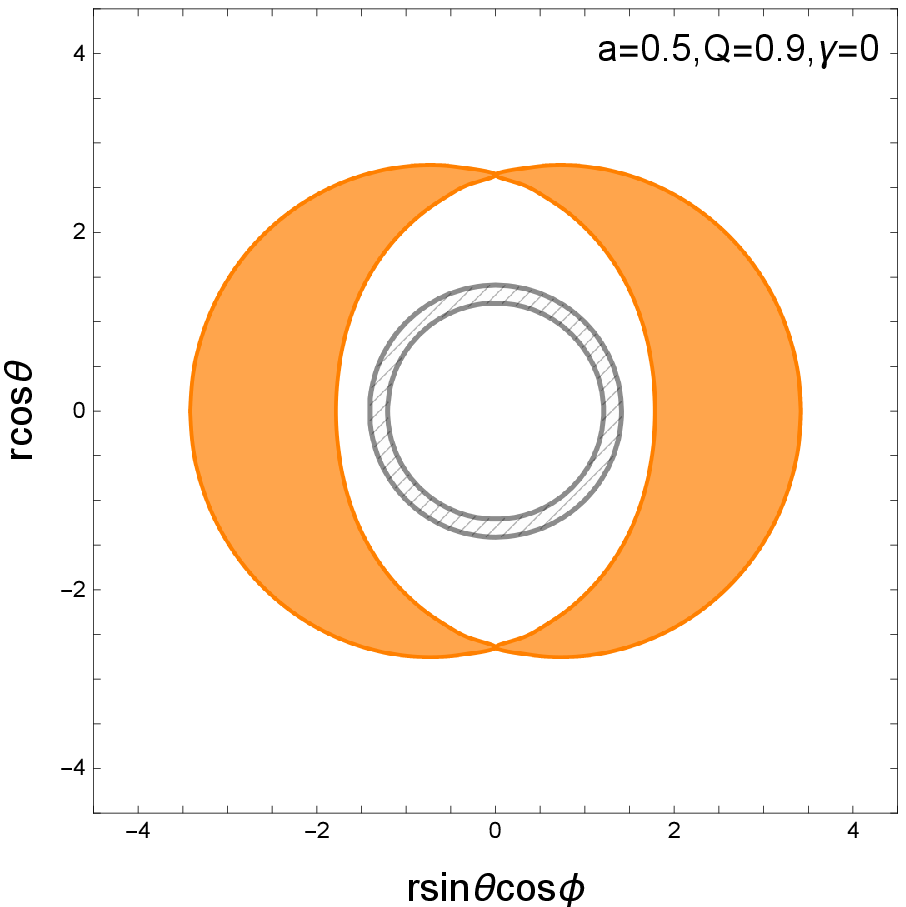}
		\vfill%
		\includegraphics[width=3.3cm,height=3.3cm]{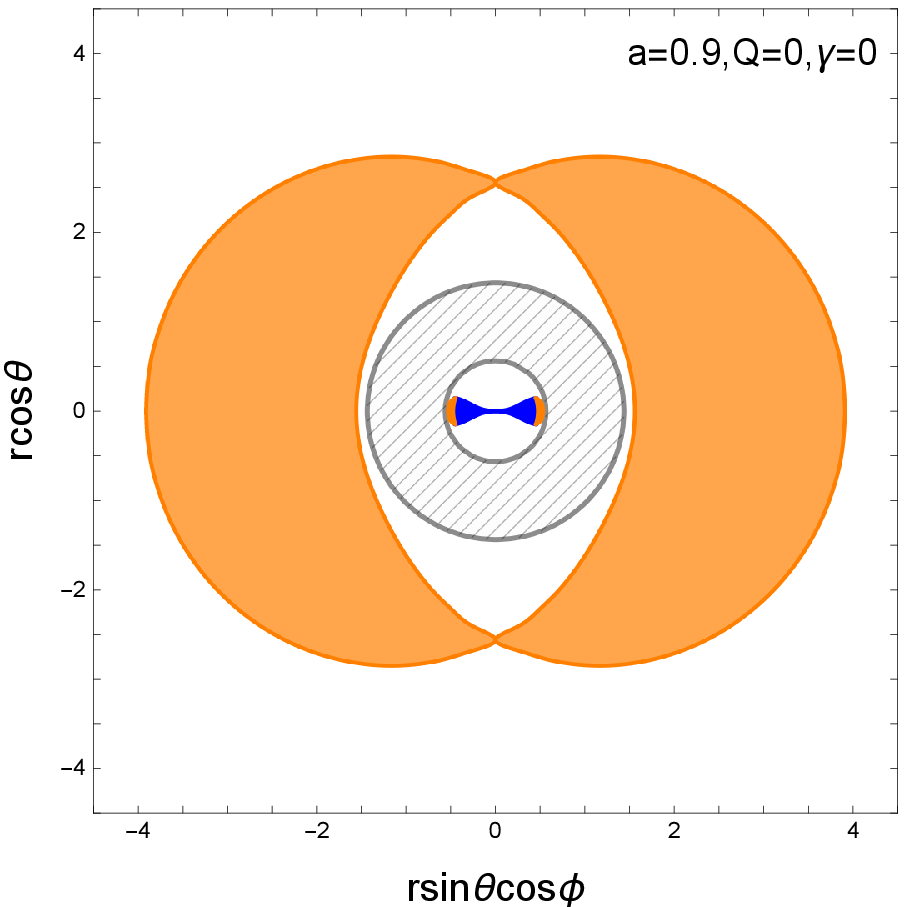}
		\hfill%
		\includegraphics[width=3.3cm,height=3.3cm]{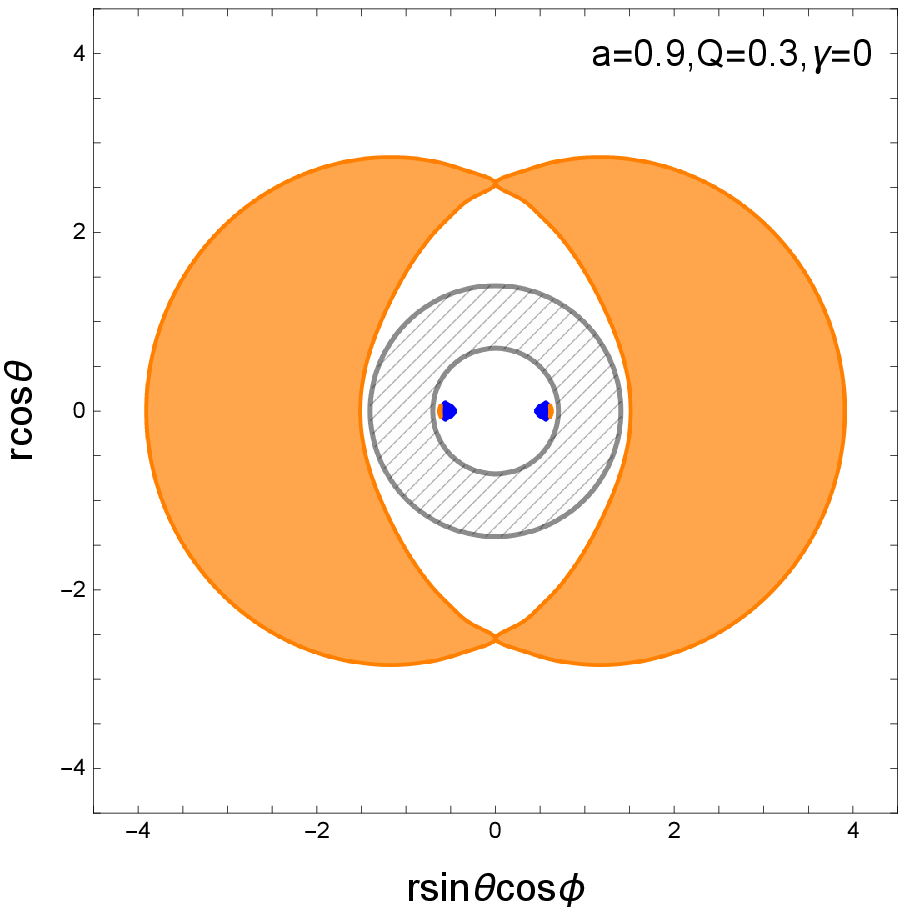}
		\hfill% 
		\includegraphics[width=3.3cm,height=3.3cm]{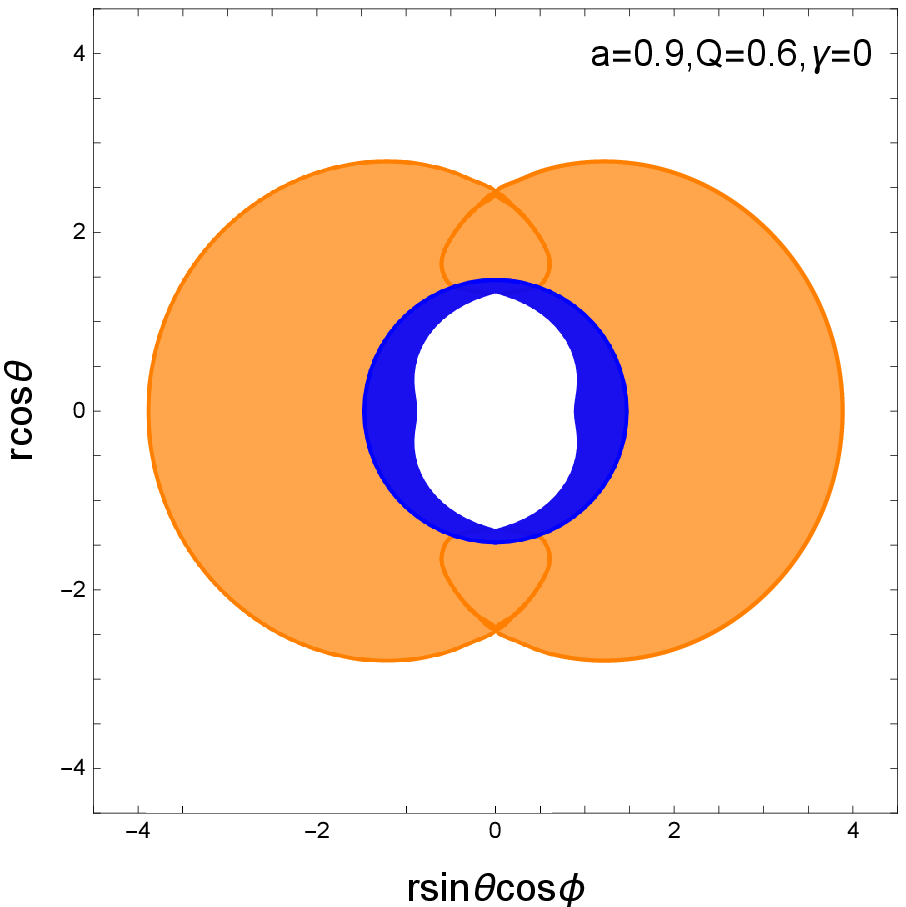}
		\hfill%
		\includegraphics[width=3.3cm,height=3.3cm]{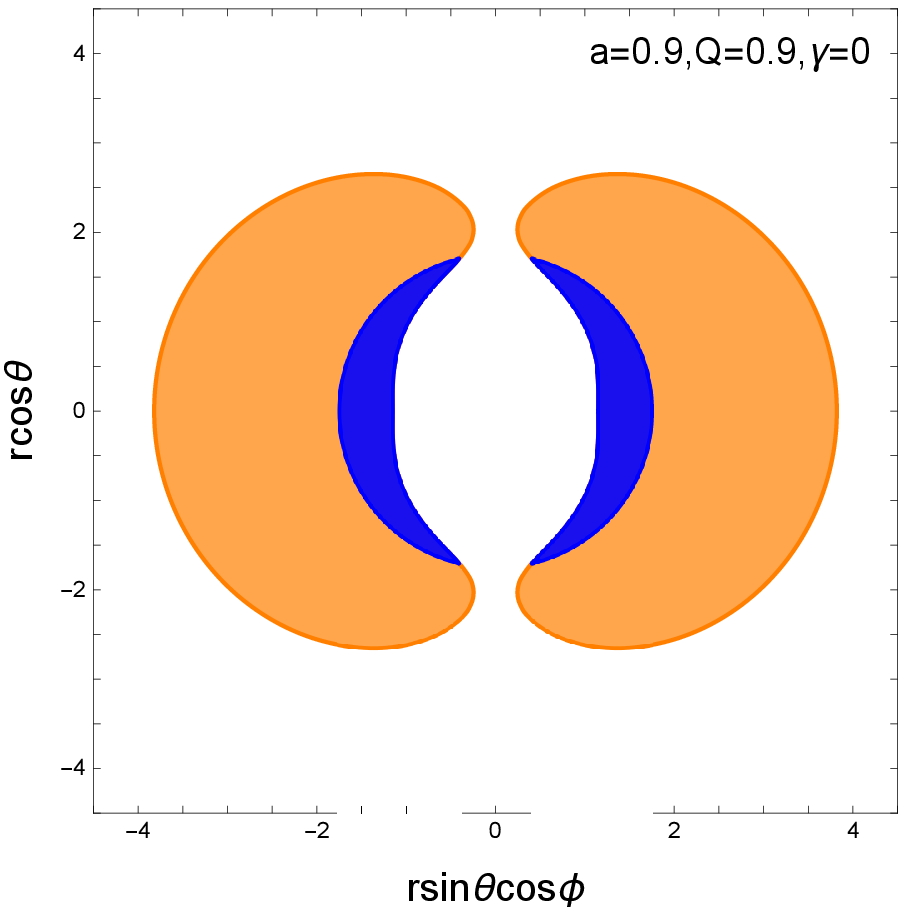}
	\end{minipage}%
	\caption{The calculated photon regions with different values of black hole parameters $a$ and $Q$. 
	The blue regions are formed by the stable orbits while the orange ones correspond to the unstable orbits.
	The shaded region satisfys $\Delta_r<0$, and therefore its boundaries $\Delta_r=0$ give rise to the outer and inner horizons.}
	\label{m2-fig-1}
\end{figure}

In Fig.~\ref{m2-fig-1}, we show the photon regions with different values of spin $a$ and charge $Q$, while fixing $\gamma=0$.
For convenience, all the quantities are expressed in units of $M$ for the remainder of the present paper.
Owing to the axisymmetry, the plots only show the intersections of the null geodesics and the plane $r\sin\theta\sin\phi =0$.
The stable and unstable orbits are distinguished by showing them in blue and orange, respectively.
The plots on the first row are for the black holes with vanishing spin $a=0$.
This is the Schwarzschild case, where the radius of the spherical orbits become degenerated, and the blue and orange regions turn into circles.
For vanishing magnetic charge $Q=0$, there is only one horizon.
As the charge increases, a second horizon, namely, the inner horizon, starts to appear from the center.
The radius of the inner horizon continues to increase as the charge increases, and it eventually merges with the outer horizon when $Q$ attains its critical value.
This feature about the horizon is similar to that of a Reissner-Nordström black hole.
Interestingly, there is a second family of null geodesics located between the two horizons.
As such null geodesics will not escape from the outer horizon, they are irrelevant to the black hole shadow in question. 
 
In the second row, the spin is taken to be a relatively small value $a=0.5$.
On the one hand, as the magnetic charge increases until its critical value $Q_c\approx0.908$, the feature of the horizon structure is again similar to that of a Reissner-Nordström black hole.
On the other hand, the characteristics of the photon region are reminiscent of that of a Kerr black hole.
Although all the null geodesics possess spherical orbits, the radii of the orbits vary between two extreme values.
In general, the motion of the photon is not circular.
To be specific, only two circular orbits are lying on the equatorial plane, attaining the extreme radii.
The smallest (largest) zenithal angle occurs for the turning point of the orbits occurs in the directions of the north (south) pole, whose radii are also unique.  
It is worth noting that there is not any photon region between the horizons.

The plots in the last row present the most intriguing features, where one has chosen a relatively large spin $a=0.9$ for the black hole metric. 
The first two plots from the left show that there are additional photon regions inside the inner horizon.
Besides unstable null geodesics, such photon region also accommodates stable null geodesics.
This is the case for either vanishing or small magnetic charges.
For a black hole with a vanishing magnetic charge, the entire area of the photon region is connected.
However, it separates into two areas as the magnetic charge starts to increase.
As the charge further increases, the two horizons coincide when the critical charge is attained, $Q = Q_c\approx0.493$.
As shown in the third plot from the left, at this moment, the two distinct areas of photon regions also become united.
One observes that the stable null geodesics are entirely attached to the unstable ones from the inside.
The last plot explore the fictitious scenario when $Q>Q_c$, the horizon disappears and, subsequently, the photon region is separated into two, while the main feature of the last plot remains.
Since the photon region inside the horizon is mostly irrelevant to the black hole shadow, we shall not process further regarding the details of its structure.

\begin{figure}[htp]
	\centering
	\begin{minipage}[c]{1\linewidth}
		\centering
		\includegraphics[width=3.5cm,height=3.5cm]{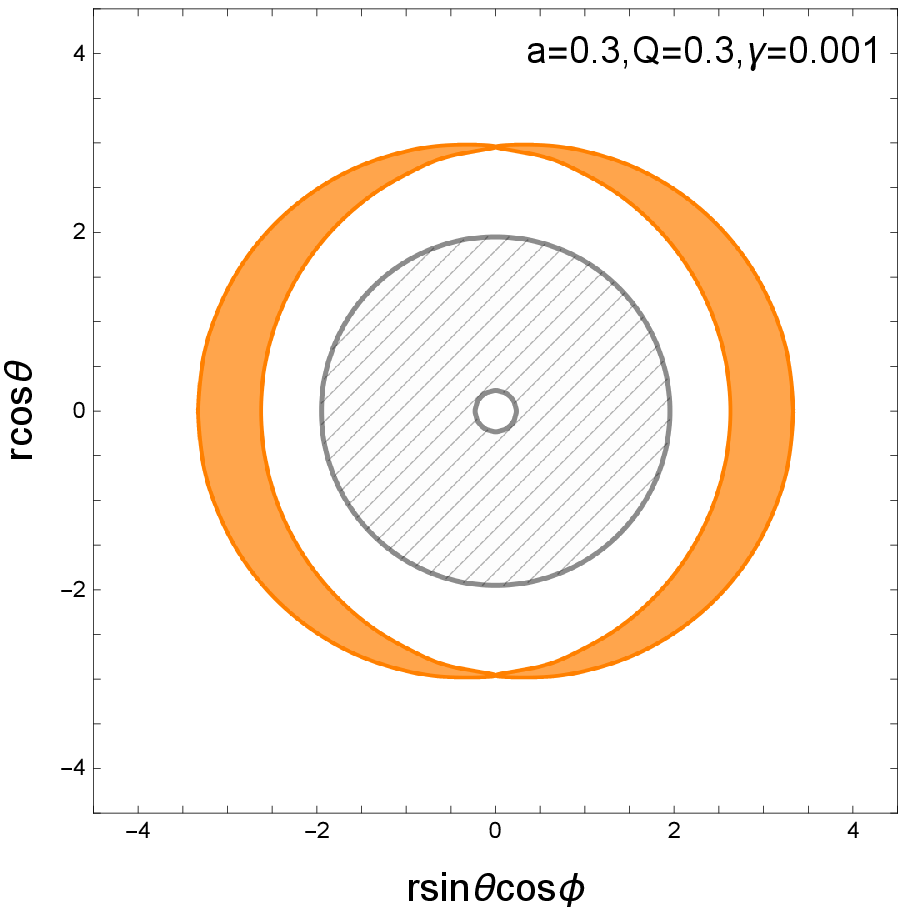}
		\includegraphics[width=3.5cm,height=3.5cm]{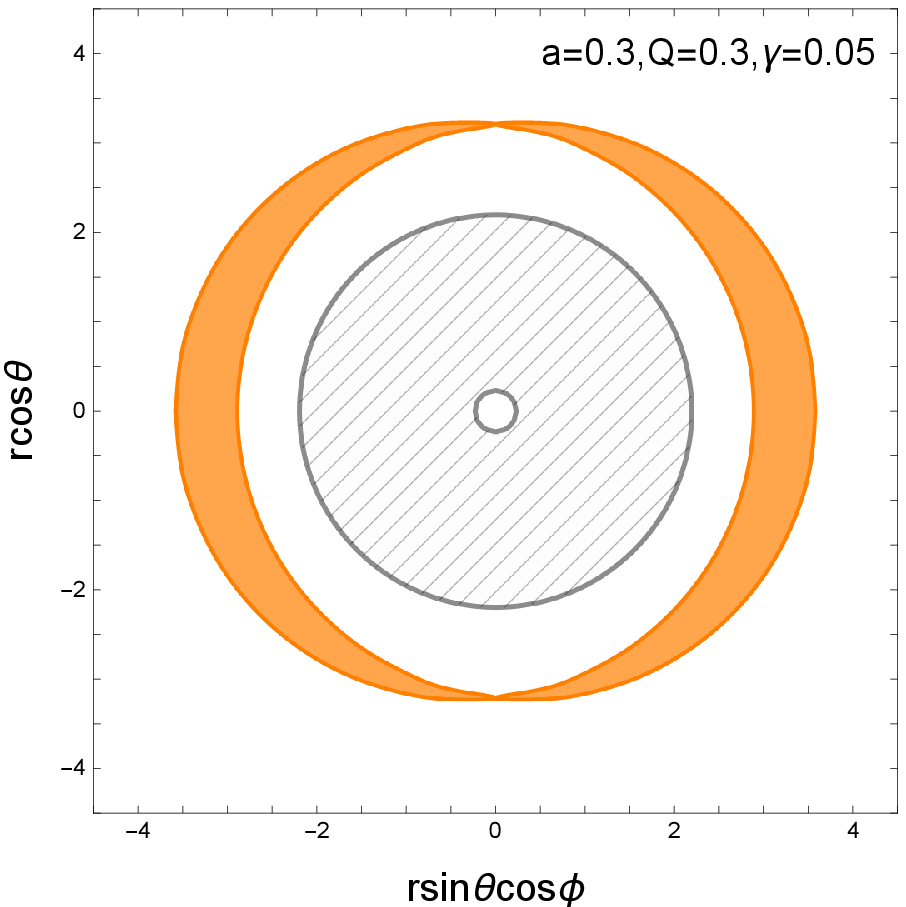}
		\includegraphics[width=3.5cm,height=3.5cm]{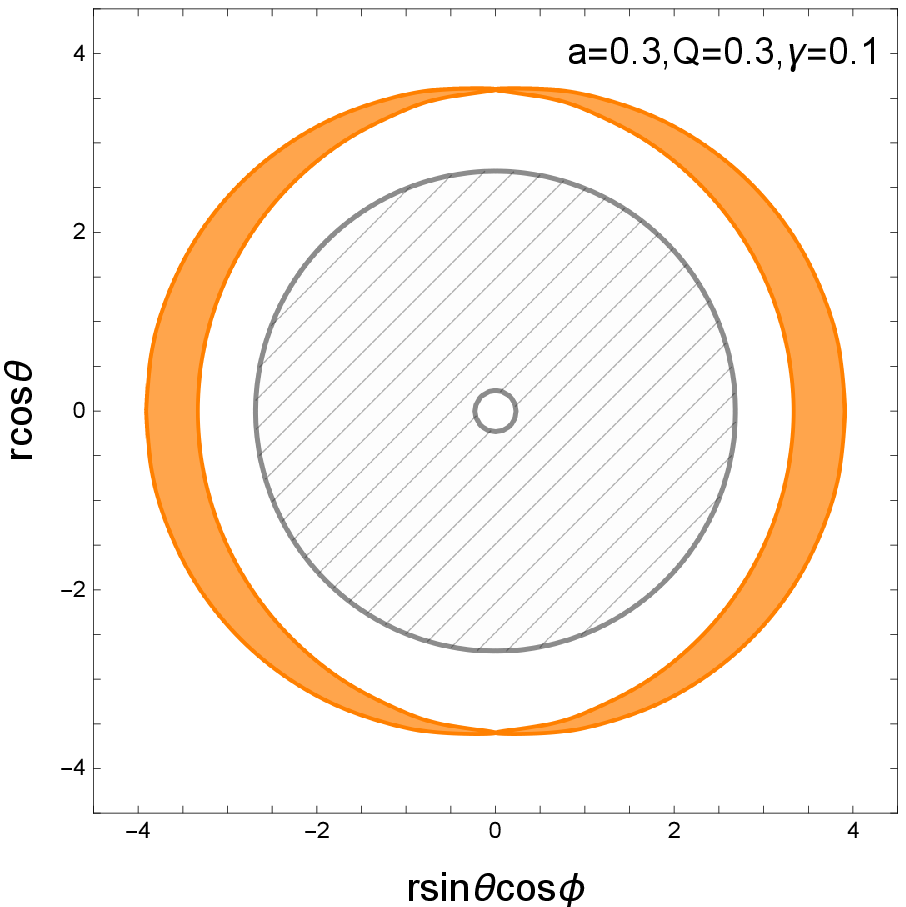}
		\includegraphics[width=3.5cm,height=3.5cm]{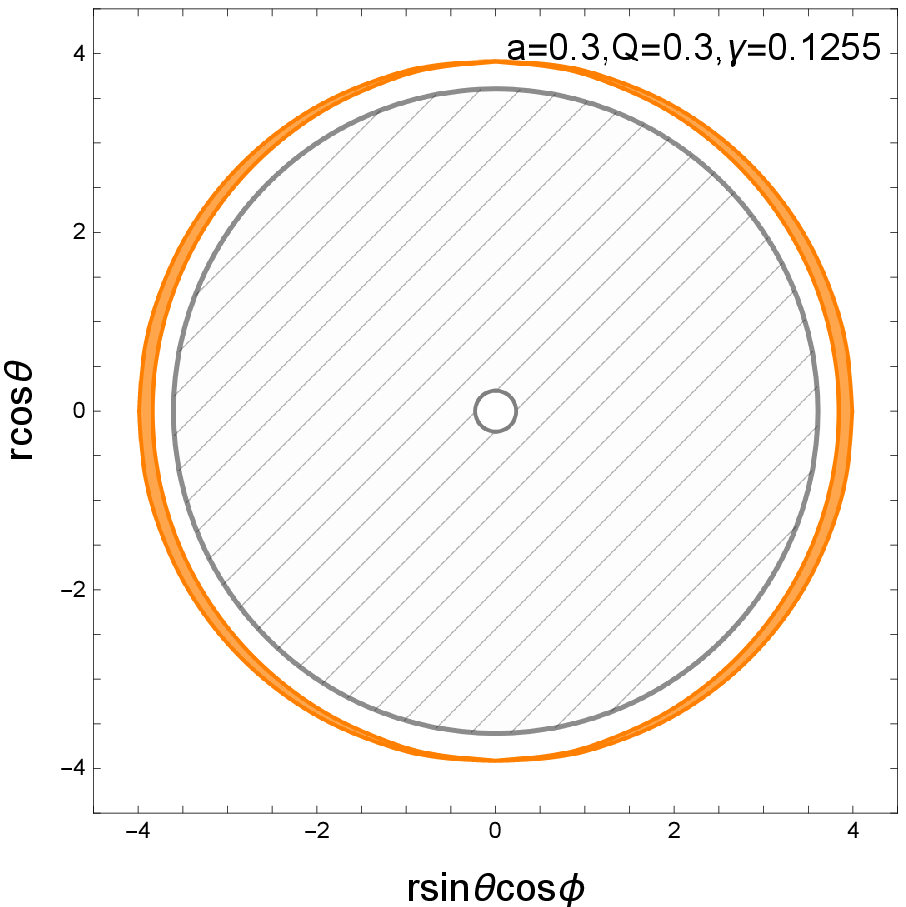}\\
		\includegraphics[width=3.5cm,height=3.5cm]{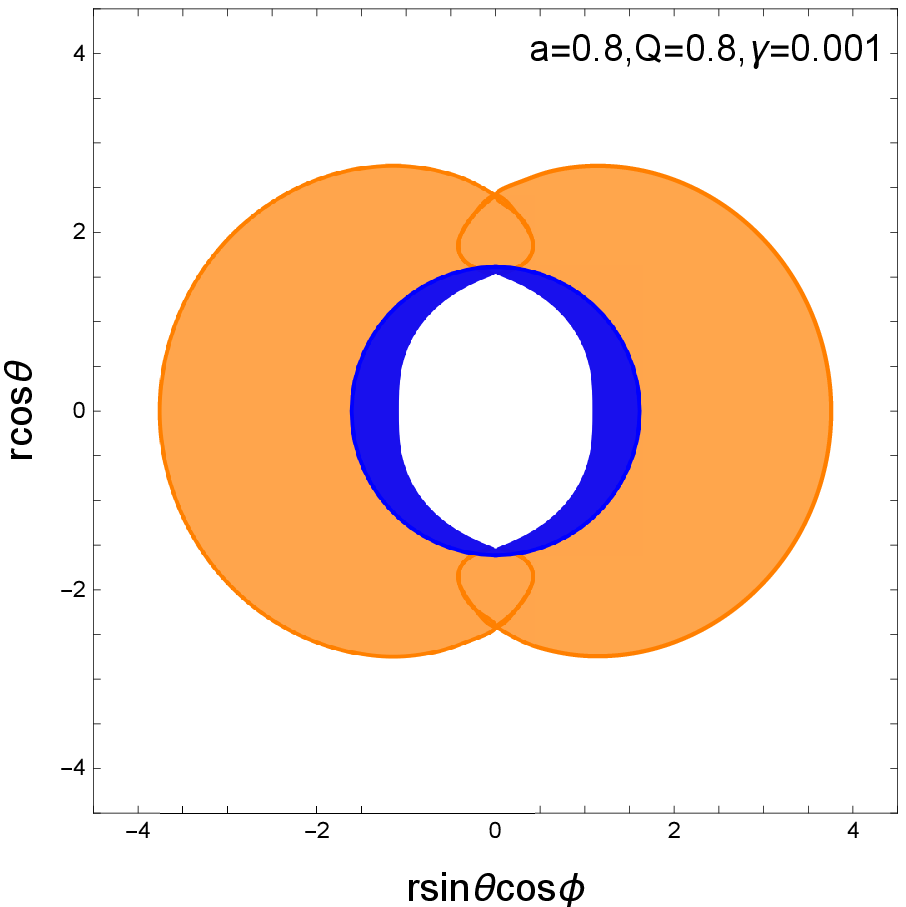}
		\includegraphics[width=3.5cm,height=3.5cm]{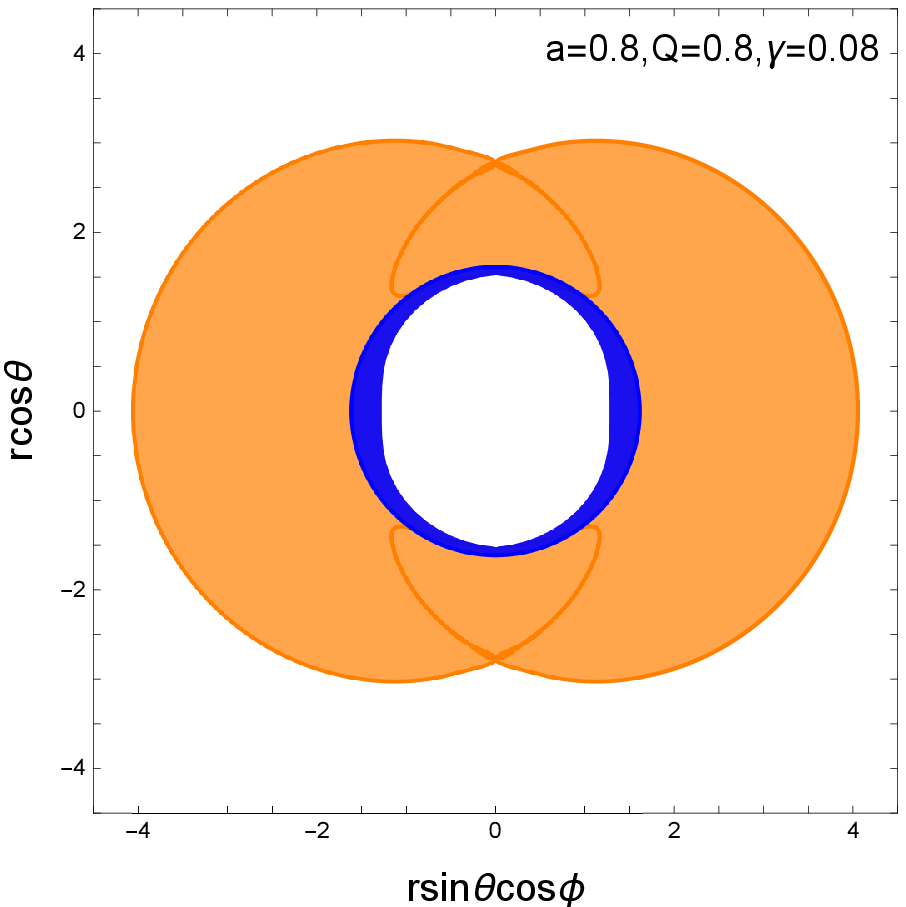}
		\includegraphics[width=3.5cm,height=3.5cm]{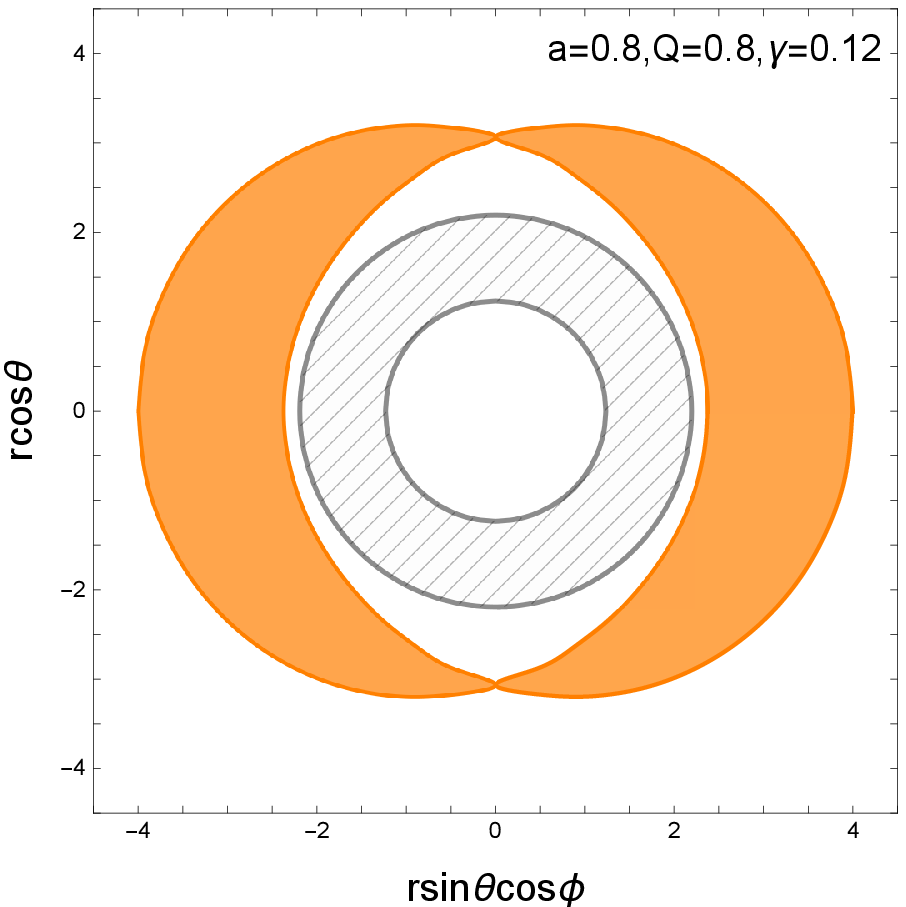}
		\includegraphics[width=3.5cm,height=3.5cm]{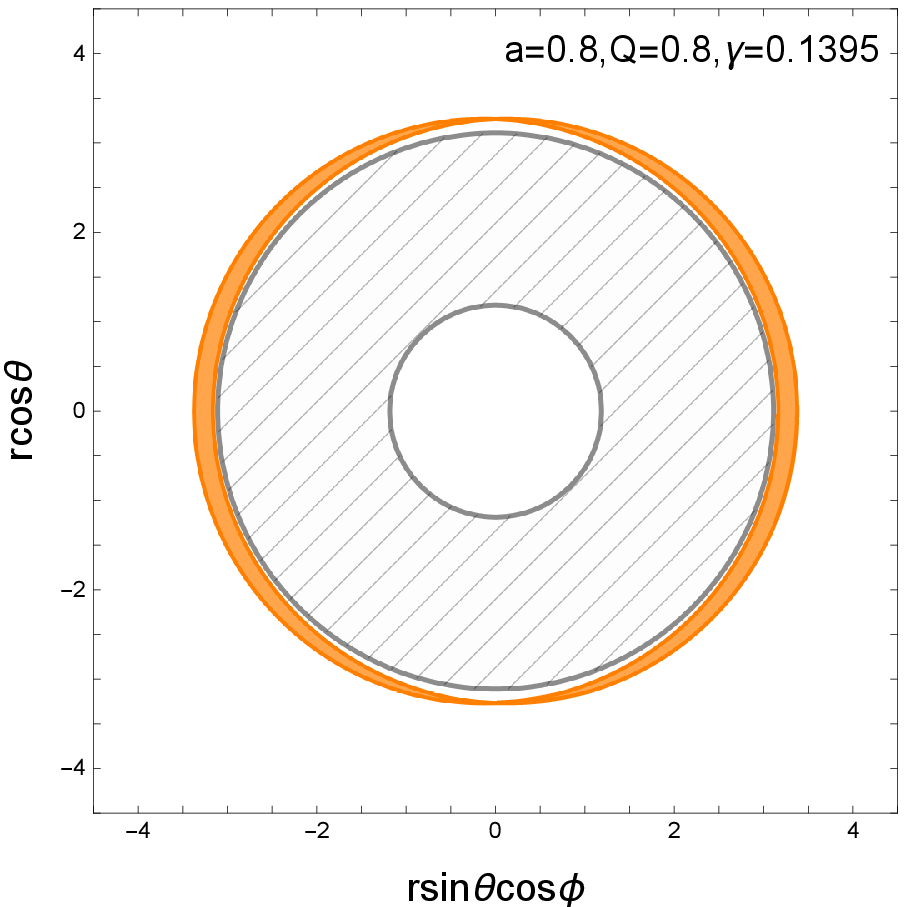}\\
		\includegraphics[width=3.5cm,height=3.5cm]{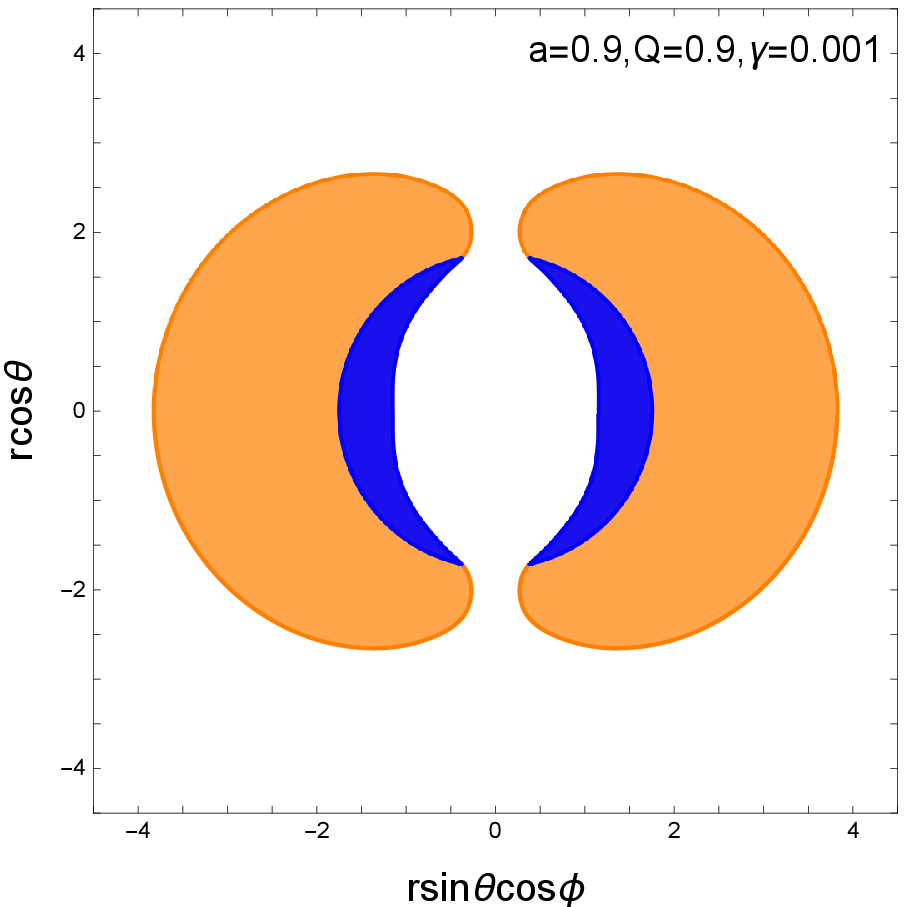}
		\includegraphics[width=3.5cm,height=3.5cm]{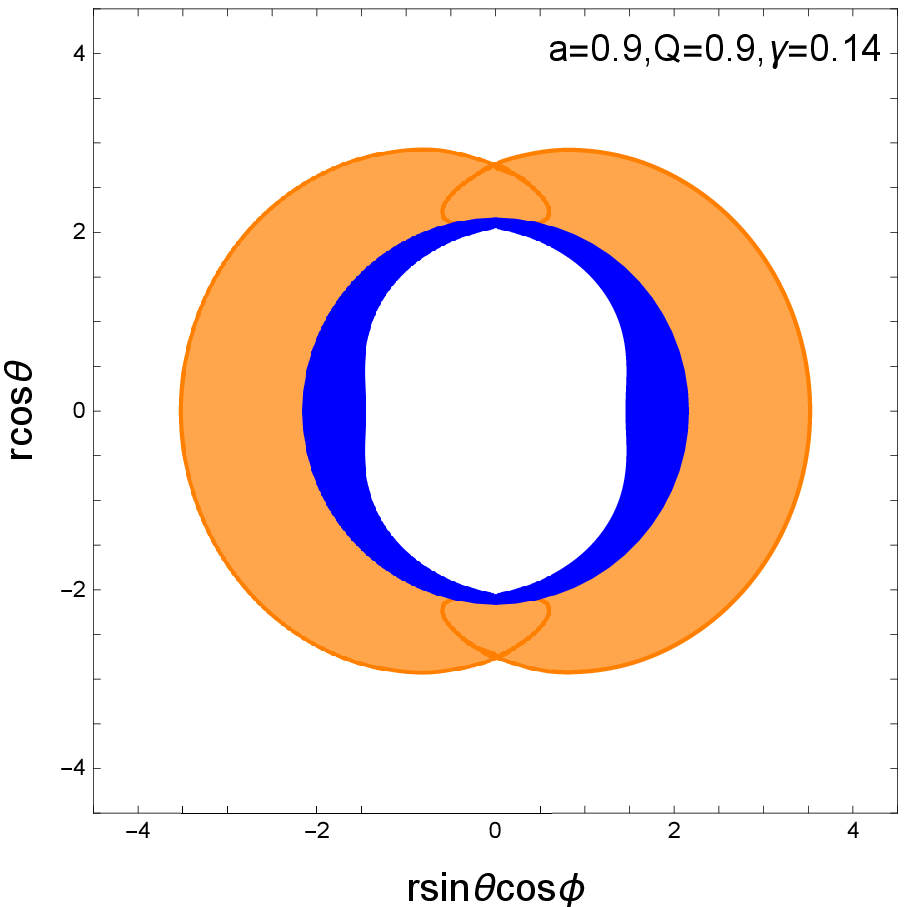}
		\includegraphics[width=3.5cm,height=3.5cm]{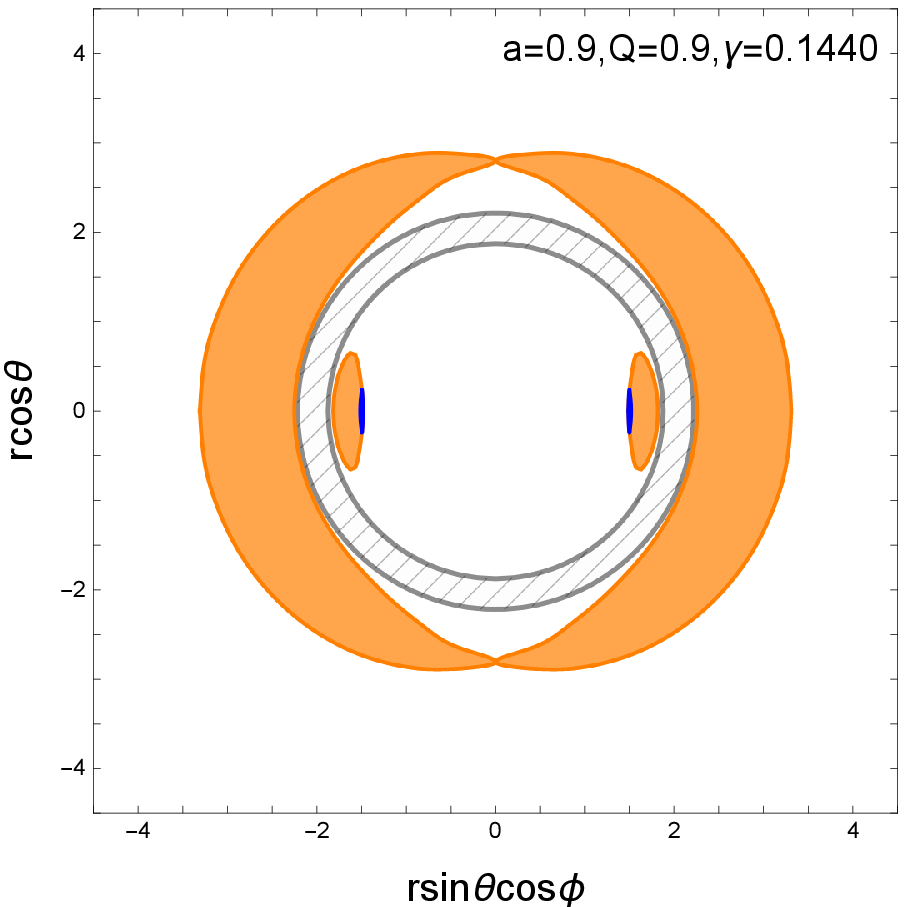}
		\includegraphics[width=3.5cm,height=3.5cm]{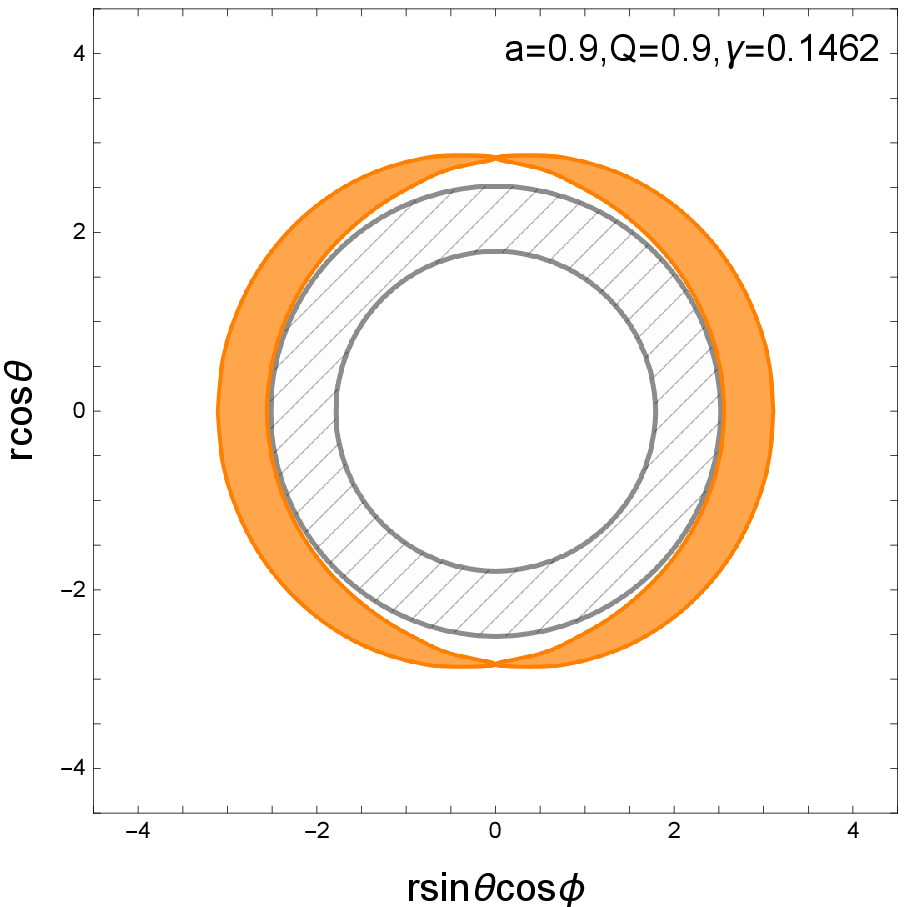}\\
		\includegraphics[width=3.5cm,height=3.5cm]{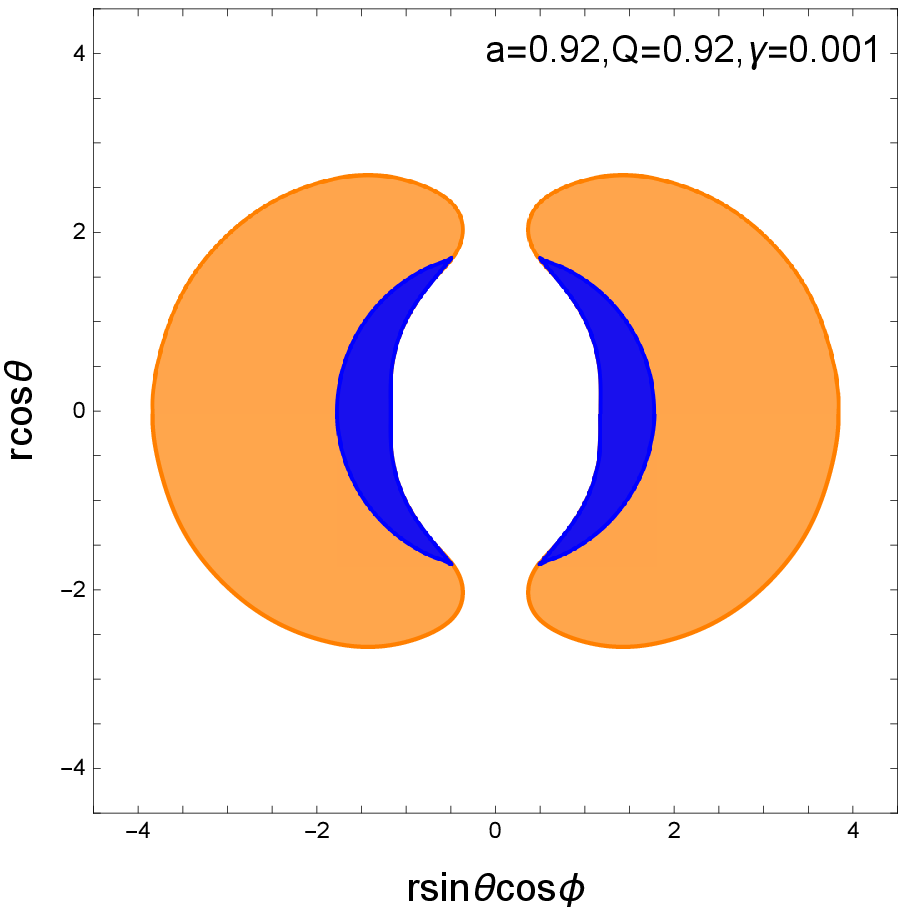}
		\includegraphics[width=3.5cm,height=3.5cm]{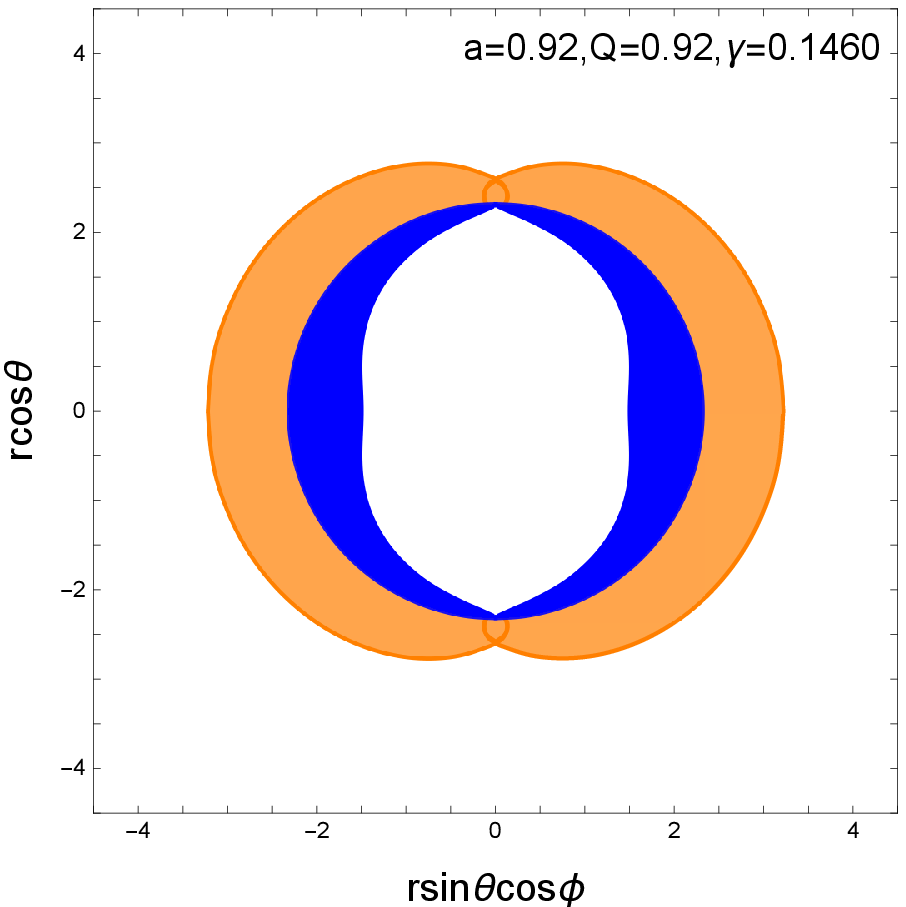}
		\includegraphics[width=3.5cm,height=3.5cm]{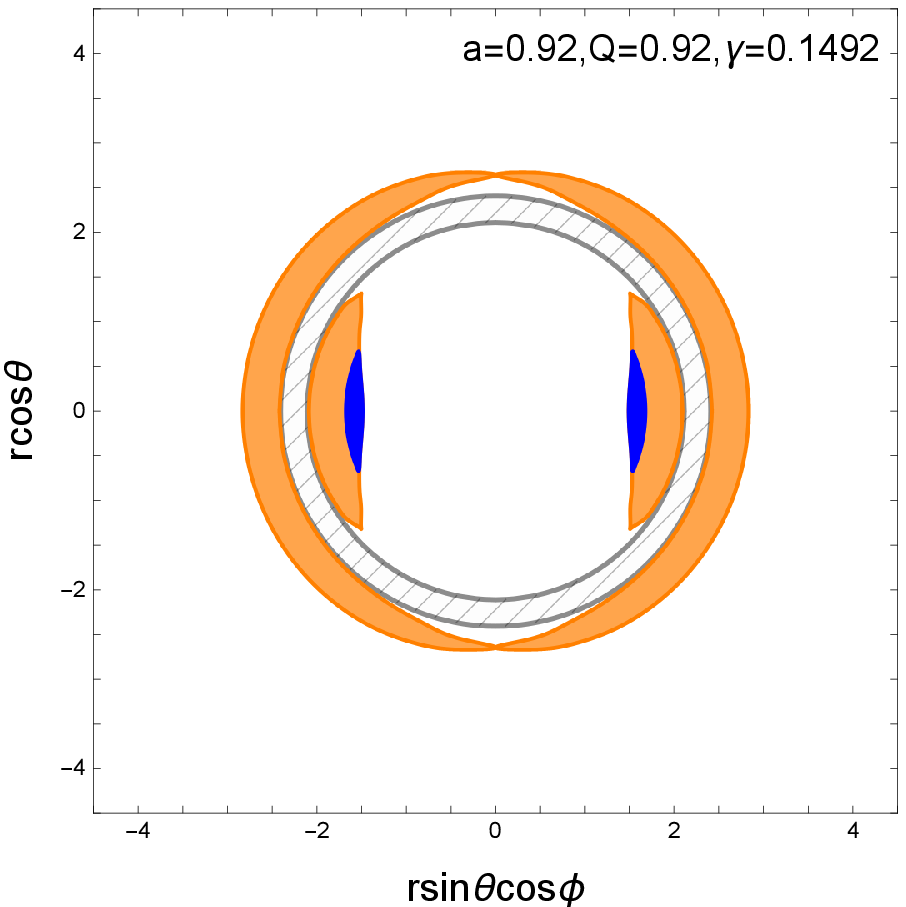}
		\includegraphics[width=3.5cm,height=3.5cm]{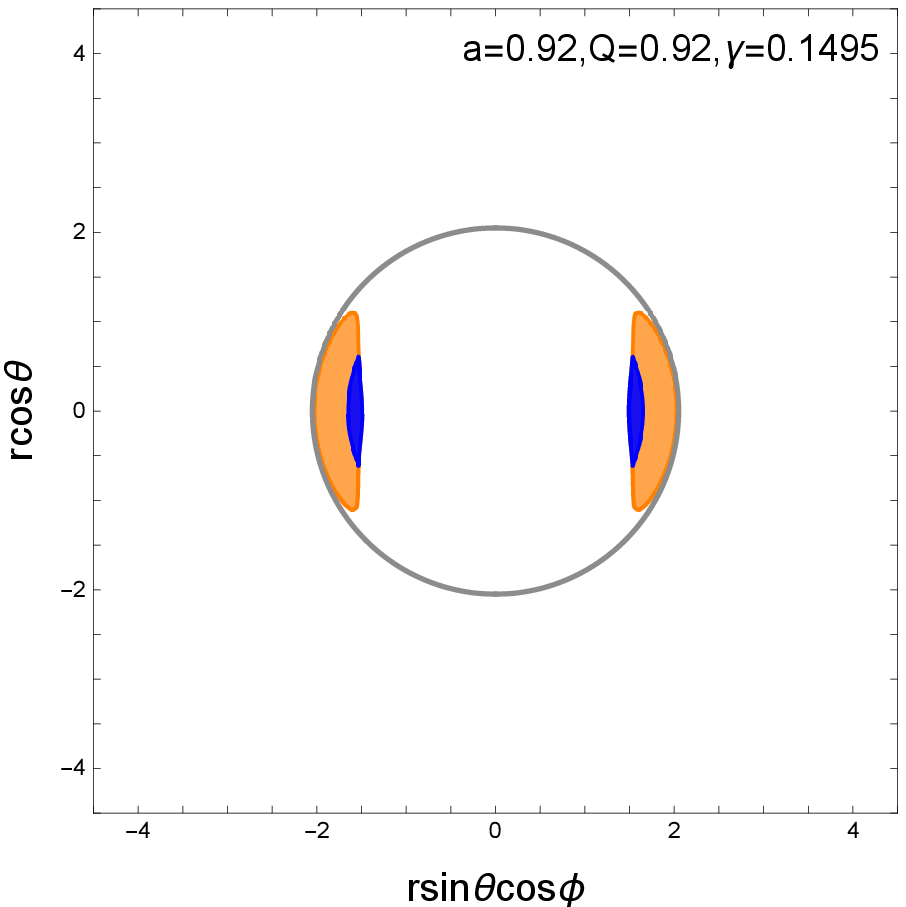}\\
		\includegraphics[width=3.5cm,height=3.5cm]{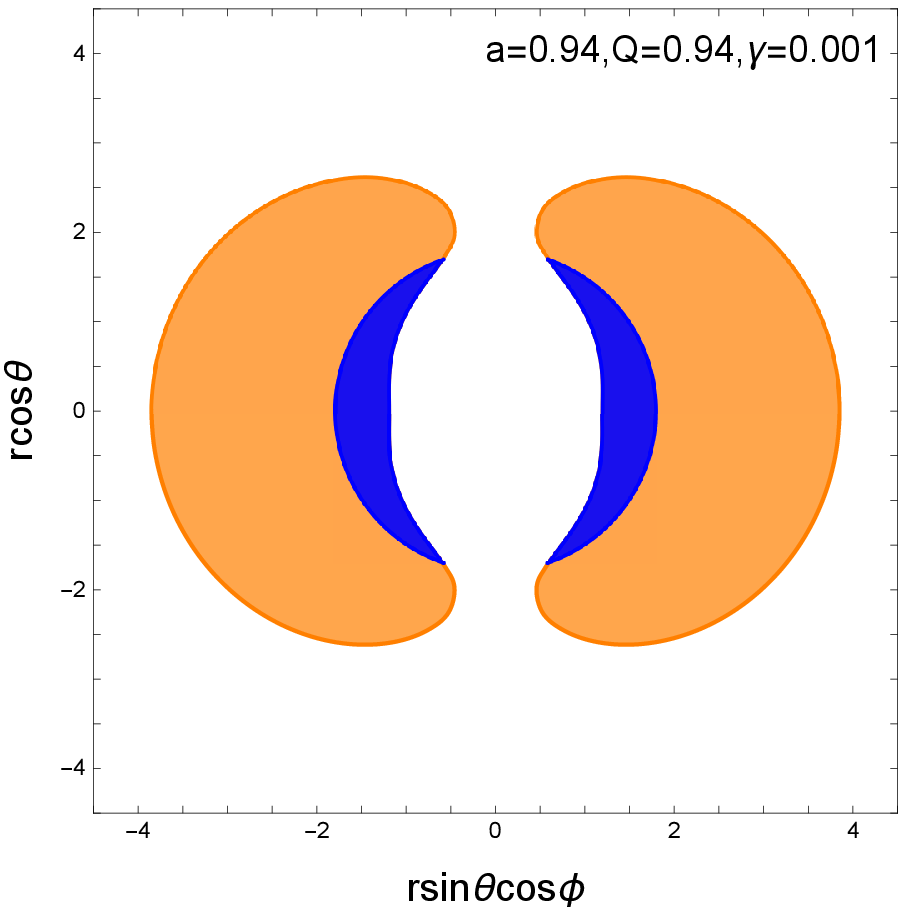}
		\includegraphics[width=3.5cm,height=3.5cm]{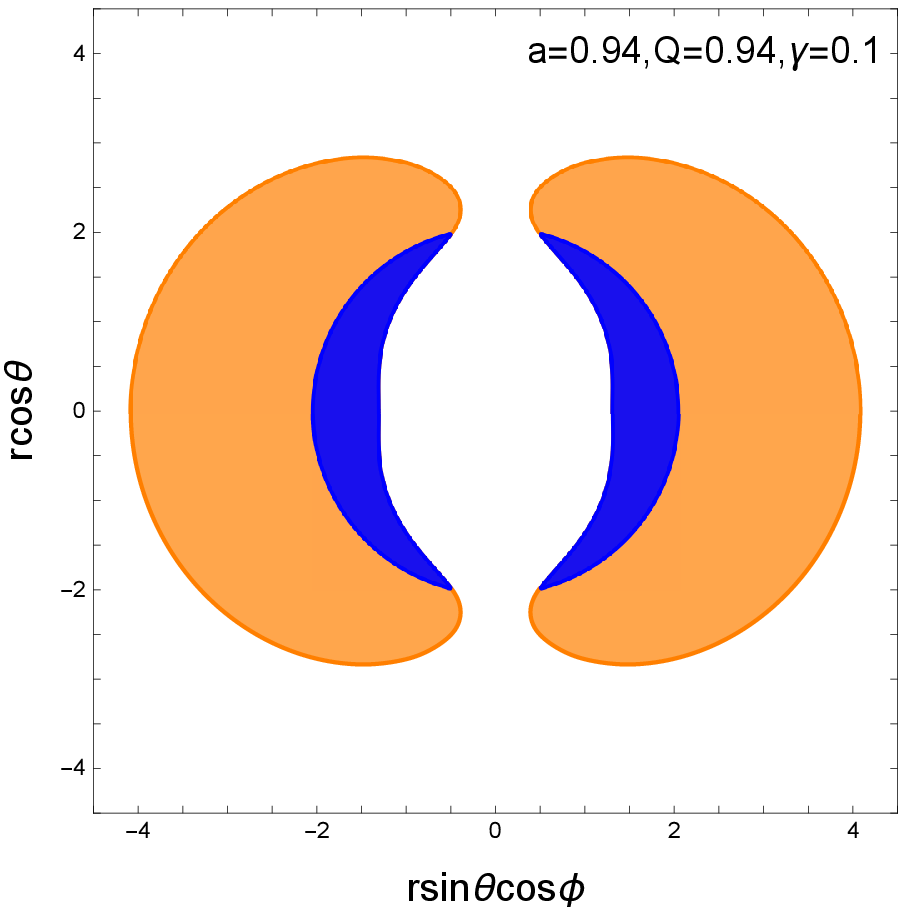}
		\includegraphics[width=3.5cm,height=3.5cm]{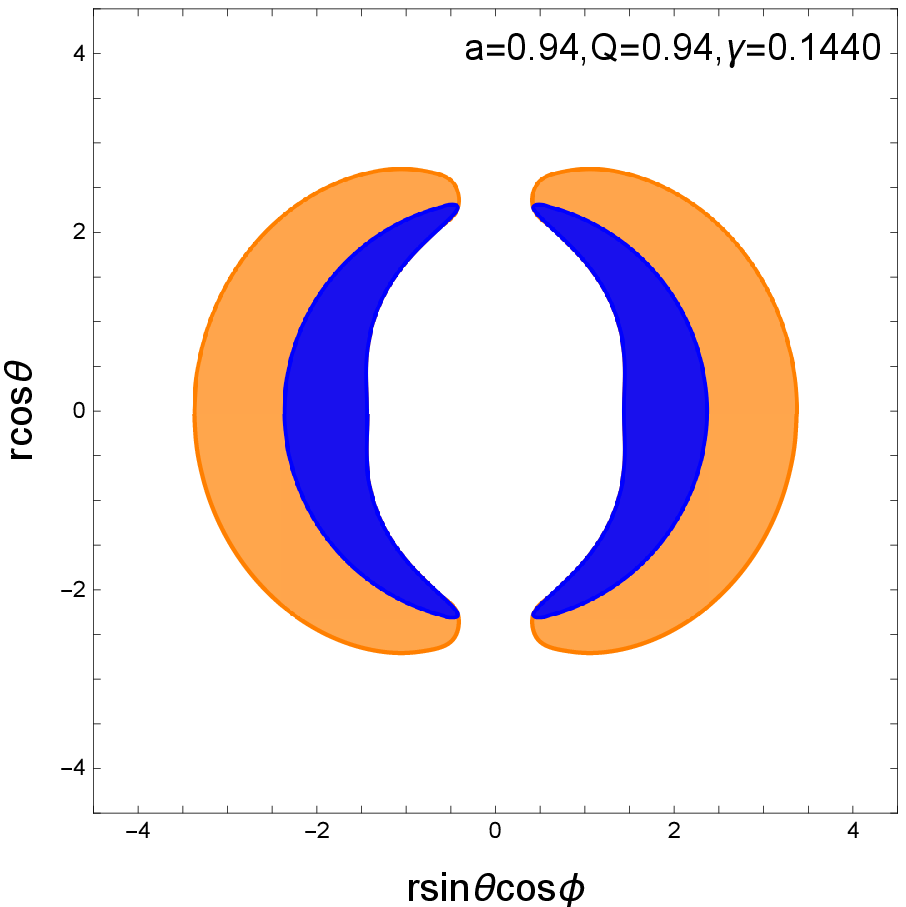}
		\includegraphics[width=3.5cm,height=3.5cm]{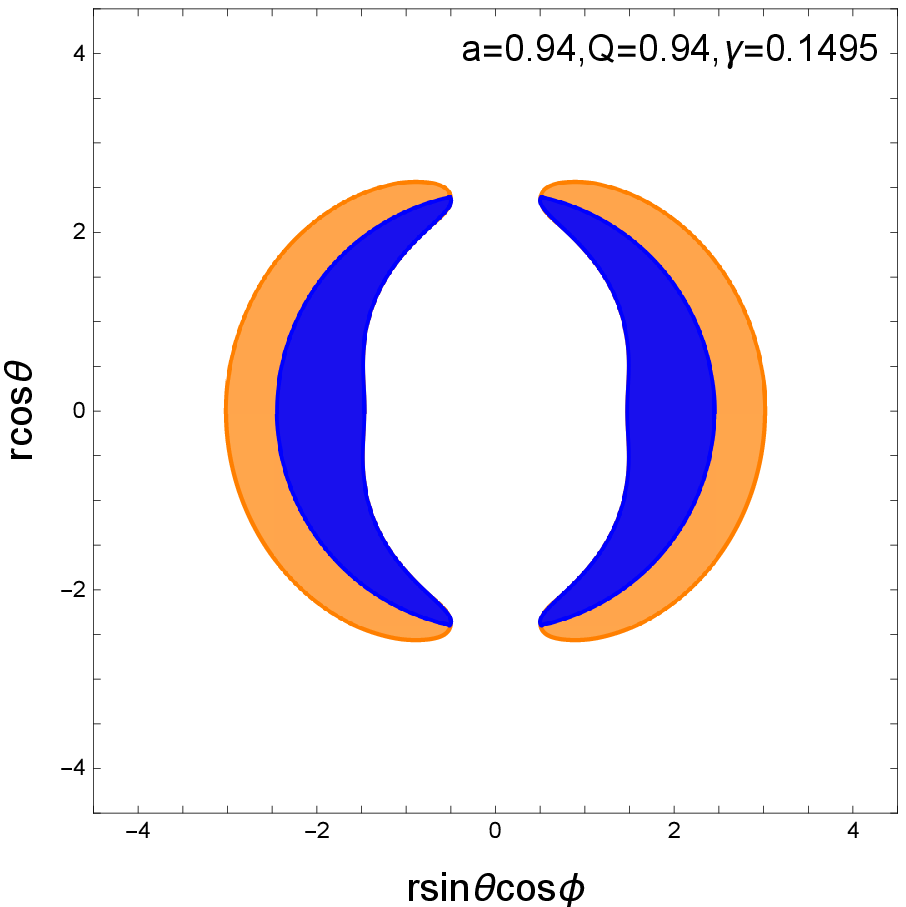}
	\end{minipage}%
	\caption{The calculated photon regions with different values of black hole parameters,  $a$ and $Q$, and $\gamma$. 
	The conventions are the same as those adopted in Fig.~\ref{m2-fig-1}, while the focus is on the role of the quintessential parameter $\gamma$.}
	\label{m2-photon2}
\end{figure}

In Fig.~\ref{m2-photon2}, we show the photon regions for different values of quintessential parameter $\gamma$. 
For each individual row, from left to right, we increase the value of $\gamma$ while assuming a given value $a=Q$.
From top to bottom, the value $a=Q$ increases from $0.3$ to $0.94$.
As observed in the plots of the first row, for small values of $a$ and $Q$, the photon regions are largely reminiscent of those of the Kerr black hole.
Also, the effect of $\gamma$ seems to compensate for that due to the spin $a$.
This is because as $\gamma$ increases, the variation of the radii of spherical orbits is suppressed.

As the values of $a$ and $Q$ increase, once again, one starts to observe intricated features of a second area of photon region inside the inner horizon, as observed in Fig.~\ref{m2-fig-1}.
The latter photon region is characterized by the appearance of stable null geodesics, which merges and gluons onto the photon region outside the horizon when the metric is in the vicinity of an extreme black hole.
Since the value of critical charge $Q_c$ decreases when $a$ increases, for a given column from top to bottom, the magnetic charge surpasses its critical value $Q=Q_c$ at a certain point.
As a result, the photon region exhibits the features which are shown in the bottom right plot of Fig.~\ref{m2-fig-1}, which are readily observed in the plots on the last row of Fig.~\ref{m2-photon2}.
On the other hand, the quintessential parameter $\gamma$ is found to largely work against the above features.
In other words, the structure tends to fall back to that of the Kerr black hole as $\gamma$ increases.
This can be observed in the rightmost plots on the second and third rows.

\section{The black hole shadows}\label{shadow}

In this section, we proceed to analyze the shadow of the black hole for an observer at spatial infinity $r=\infty$. 
In order to investigate the deformation of black hole shadow with respect to various physical parameters, we employ two different approaches.

In asymptotically flat spacetime, the size of the black hole shadow can be measured by an observer at infinity $r=\infty$ in terms of the celestial coordinates $\alpha$ and $\beta$~\cite{Bardeen1973}.
\begin{eqnarray}
&&\alpha=\lim_{r\rightarrow \infty}\left(-r\sin\theta\frac{d\phi}{dr}\mid_{(r_0,\theta_0)}\right),\\
&&\beta=\lim_{r\rightarrow \infty}\left( r^2\frac{d\theta}{dr}\mid_{(r_0,\theta_0)}\right),
\end{eqnarray}
where $r_0$ represents the distance from the black hole to the observer, and $\theta_0$ is the inclination angle between the line of observer's sight and the spinning axis of the black hole.

The above limit can be evaluated by using the specific form of the metric, and expressed in terms of the constants of motion $\xi$ and $\eta$ given in Eqs.~(\ref{equ3}) and (\ref{equ4}) as follows
\begin{eqnarray}
&&\alpha=-\frac{\xi}{\sin{\theta_0}},\\
&&\beta=\pm \sqrt{\eta-(\xi\csc{\theta_0}-a\sin\theta_0)^2},
\end{eqnarray}
For further simplification, one assumes that the observer is located on the equatorial plane of the black hole, thus the inclination angle $\theta_0=\frac{\pi}{2}$. One has
\begin{eqnarray}
&&\alpha=-\xi,\\
&&\beta=\pm \sqrt{\eta-(\xi-a)^2},
\end{eqnarray}

 \begin{figure}[htp]
	\centering
	\begin{minipage}[c]{0.7\linewidth}
		\centering
		\includegraphics[width=4.2cm]{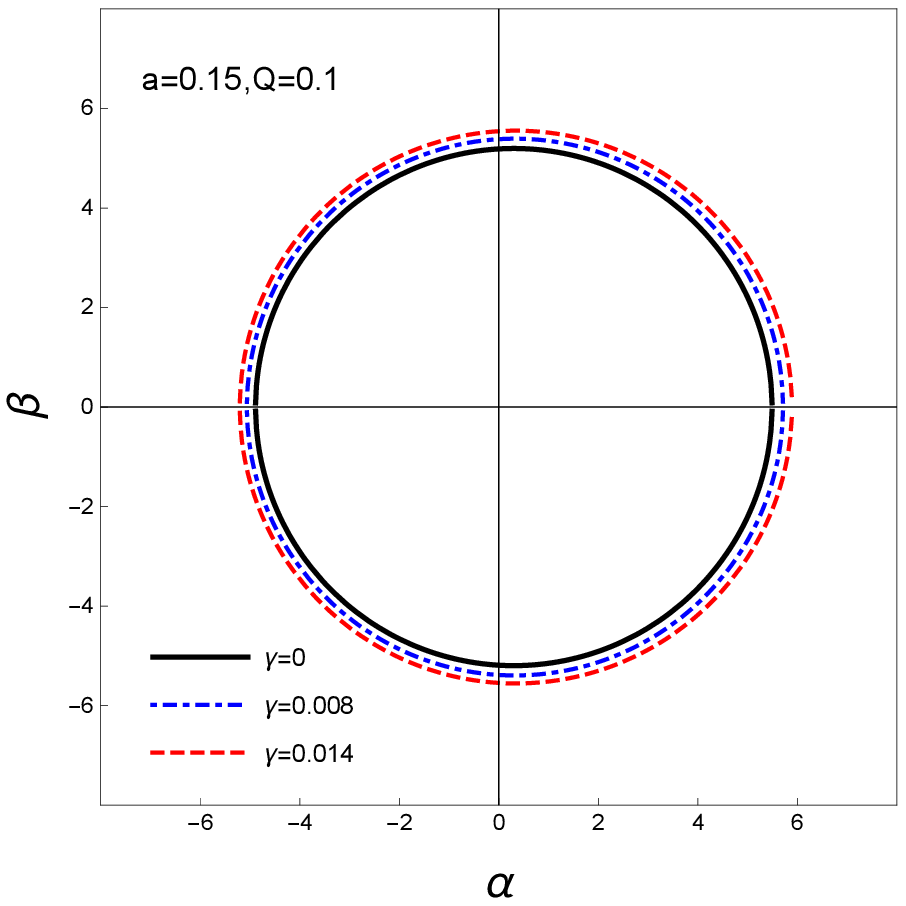}
		\includegraphics[width=4.2cm]{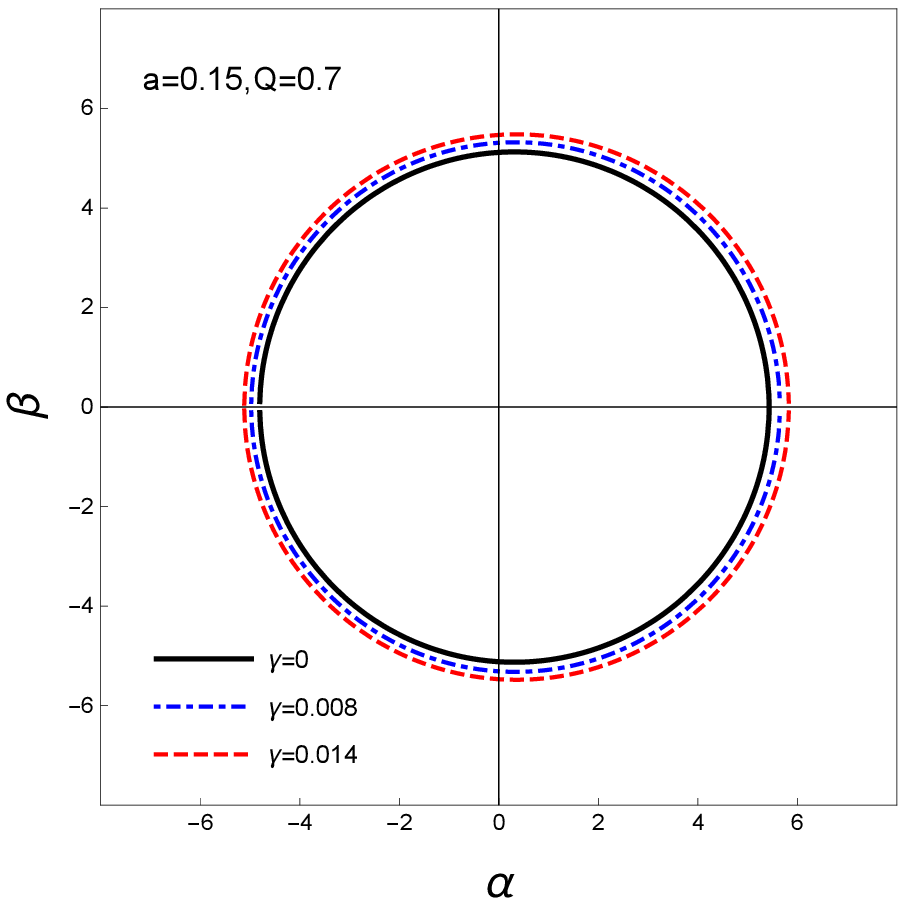}
		\includegraphics[width=4.2cm]{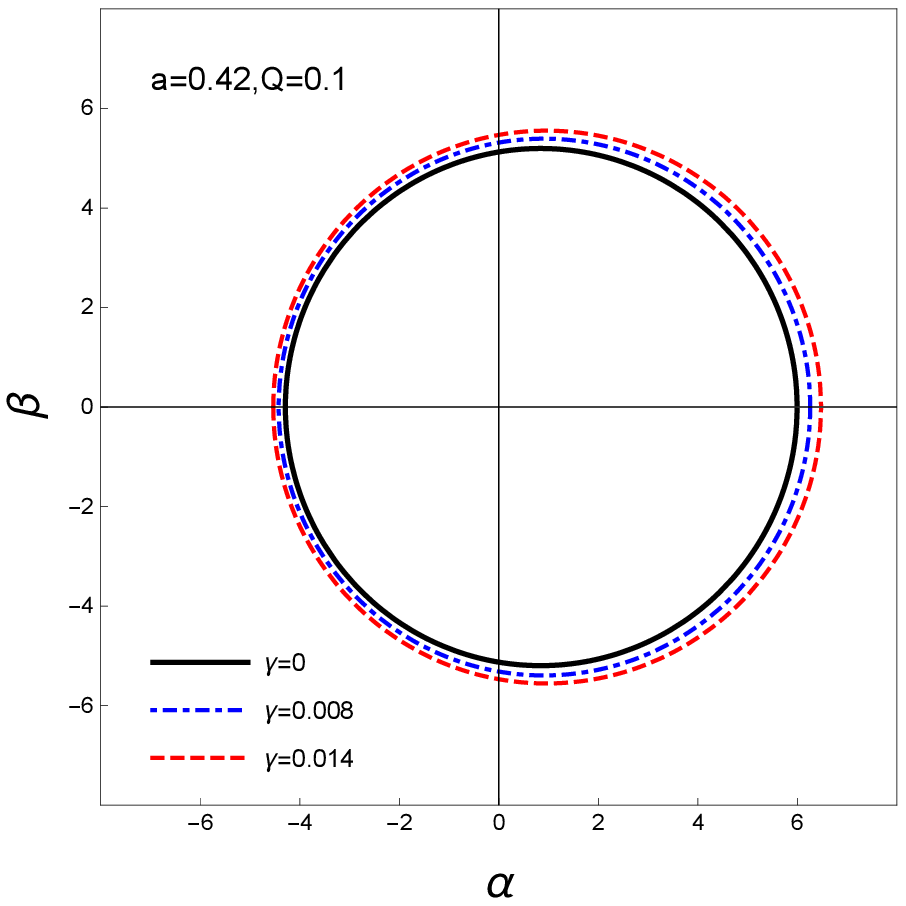}
		\includegraphics[width=4.2cm]{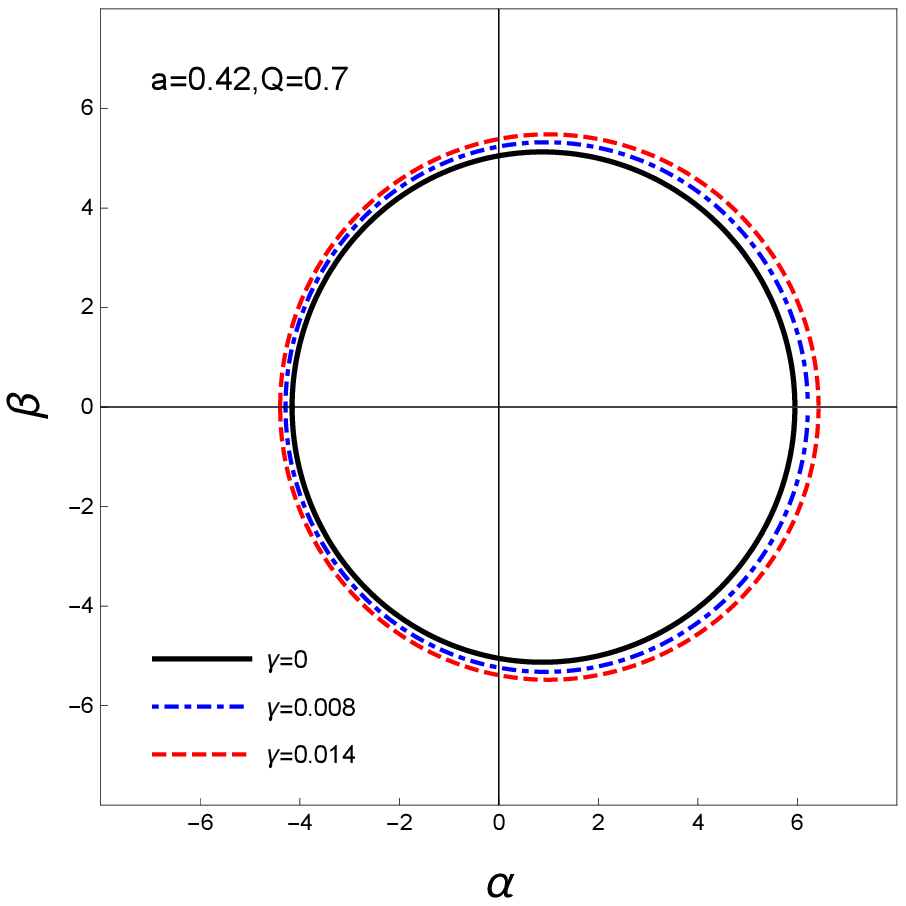}
		\includegraphics[width=4.2cm]{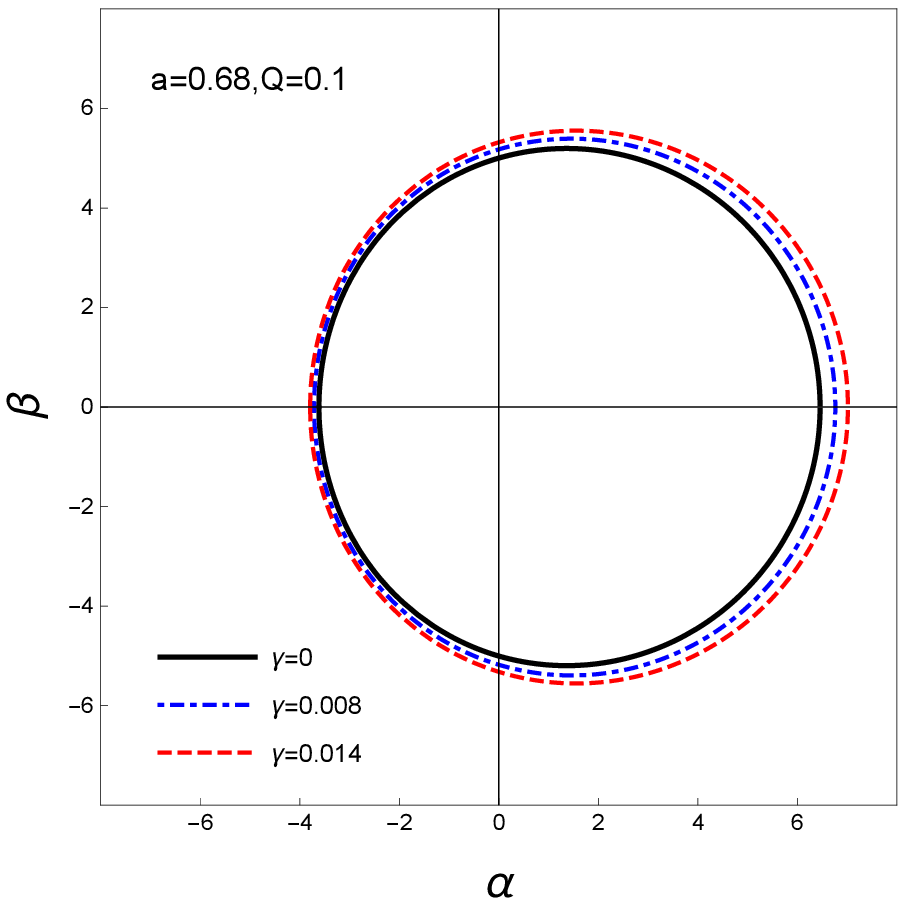}
		\includegraphics[width=4.2cm]{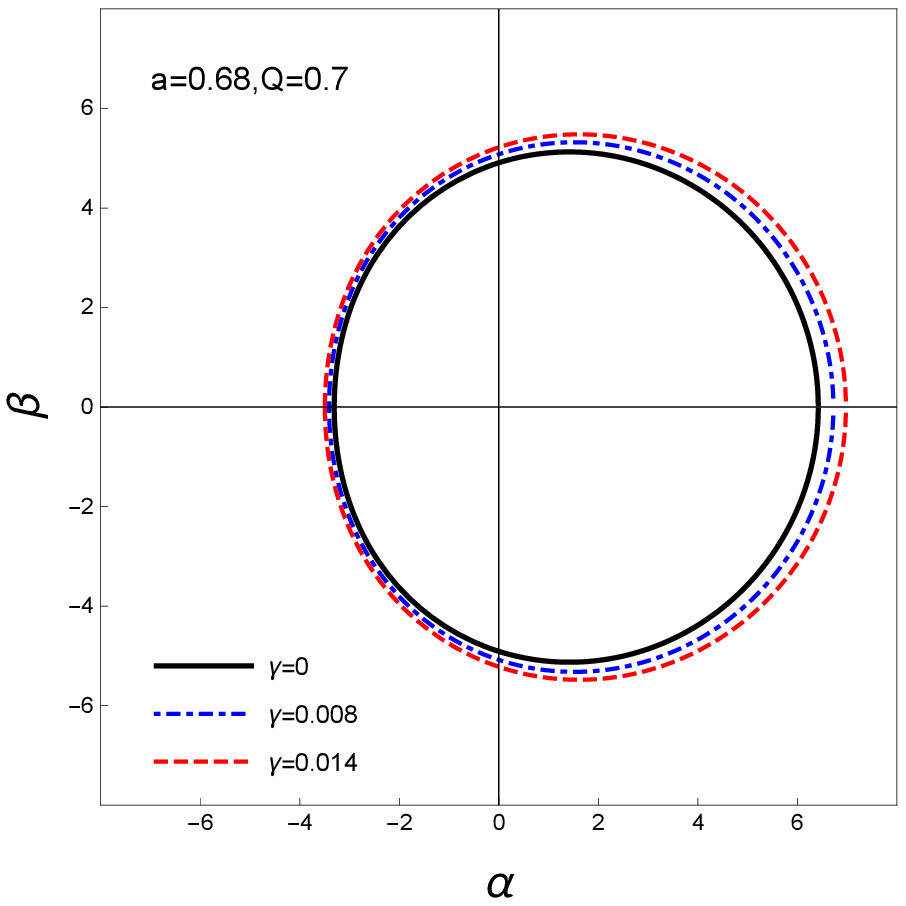}
	\end{minipage}%
	\caption{The calculated black hole shadows for different metric parameters $a$, $Q$, and $\gamma$.
	The observer is located at spatial infinity.}
	\label{model2-shad}
\end{figure}

In Fig.~\ref{model2-shad}, we show the calculated black hole shadow for different metric parameters $a$, $Q$, and $\gamma$. 
From the two plots in the first row, one observes that for small black hole spin $a=0.15$, the charge $Q$ and quintessential parameter $\gamma$ do not significantly affect the shape of the shadow.
But they do have an impact on the overall size of the shadow.
As one increases the black hole spin in the second and third rows, the shadow is gradually deformed, as it becomes more asymmetric with respect to the vertical axis $\alpha=0$.
On the other hand, as $\gamma$ increases, the size of the black hole shadow increases, while the deformation also becomes slightly more significant.
Since the above modifications are not significant, in the next section we will employ a few quantities to quantify the deformation. 

 \begin{figure}[htb]
	\centering
	\begin{minipage}[c]{1\linewidth}
		\centering
		\includegraphics[width=4cm]{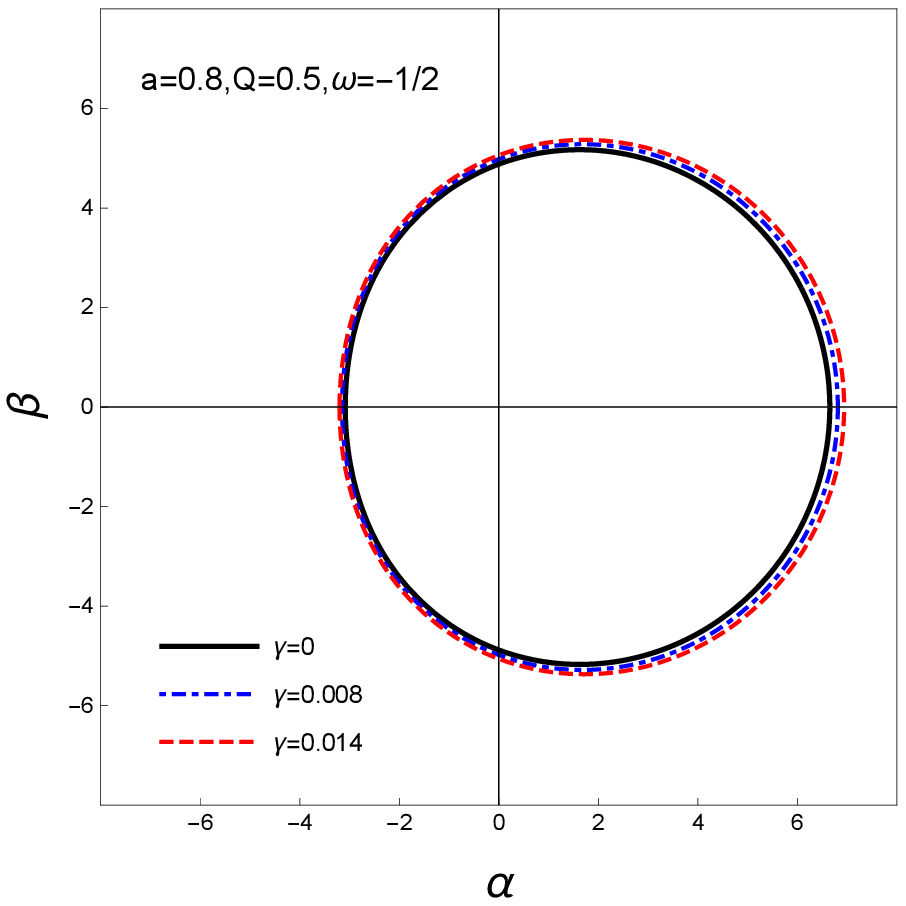}
		\includegraphics[width=4cm]{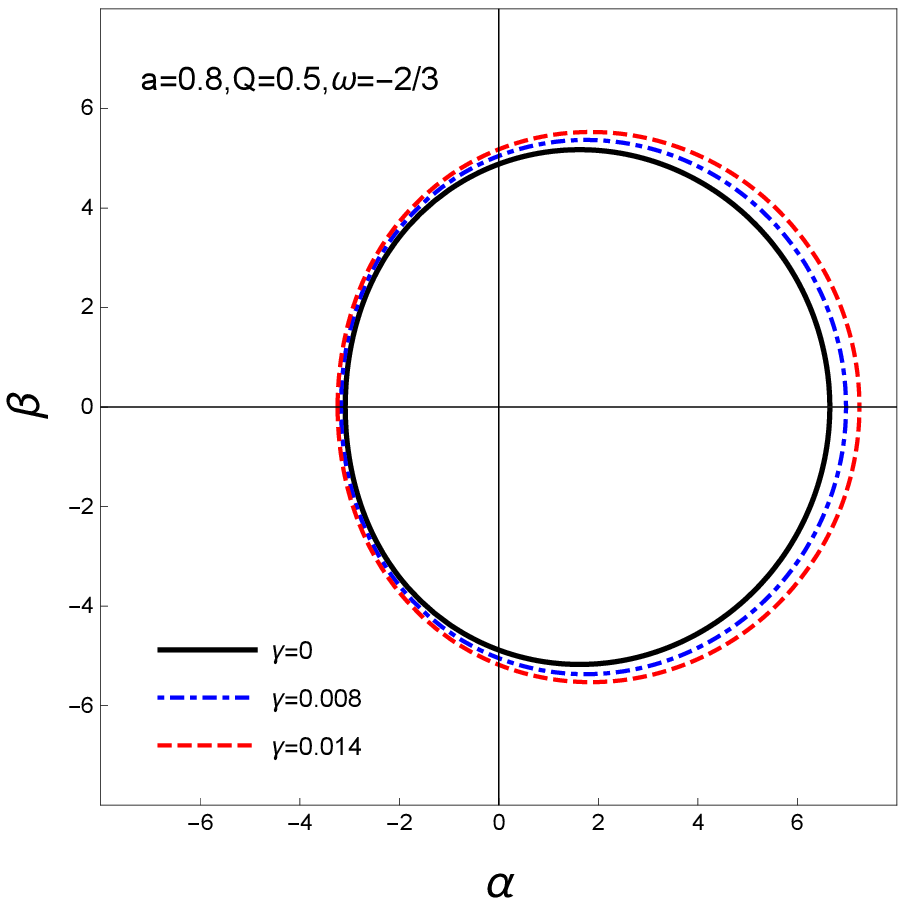}
		\includegraphics[width=4cm]{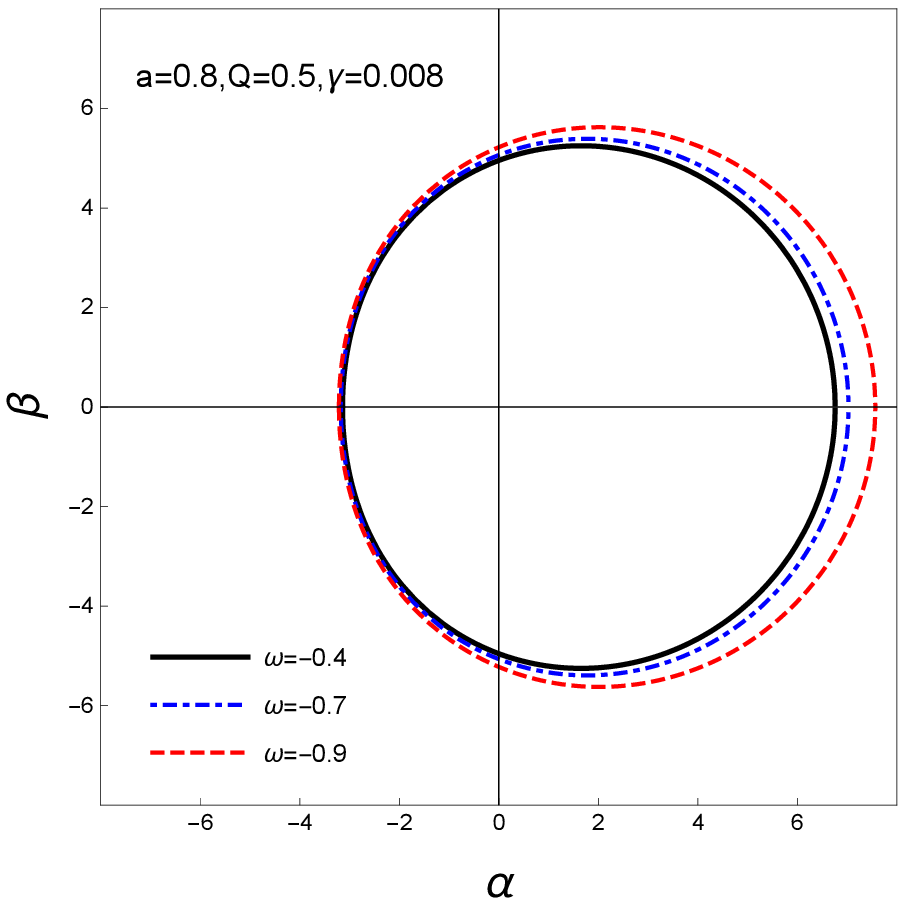}
	\end{minipage}%
	\caption{The calculated black hole shadows for different state parameter $\omega$.
	The observer is located at $r_0=50$.}
	\label{model2-ome}
\end{figure}

We also investigate the effect of the state parameter $\omega$ on the black hole shadow.
The results are shown in Fig.~\ref{model2-ome}.
The left plot is calculated by taking $\omega=-1/2$, while for the middle one, we assume $\omega=-2/3$.
Apparently, the distortion depends more sensitively on $\gamma$ in the case of $\omega=-2/3$.
The right plot shows the dependence of the shadow on $\omega$, by fixing $a=0.8, Q=0.5$, and $\gamma=0.008$. 
It is observed that as $\omega$ decreases, the area of the black hole shadow increases.
We will analyze the results in a more quantitative fashion in the next section.

\subsection{Quantitative observables and the energy emission rate}\label{emission}

In order to analyze the apparent shape of the shadow in detail, we adopt the two observables introduced in~\cite{Hioki2009}, namely, the shadow radius $R_s$ and the distortion parameter $\delta_s$.
Here, $R_s$ is defined as the radius of reference circle that passes through the three points $T(\alpha_t,\beta_t)$ the top one, $B(\alpha_b,\beta_b)$ the bottom one, and $R(\alpha_r,\beta_r)$ the right one, by the following expression
\begin{eqnarray}
R_s=\frac{(\alpha_t-\alpha_r)^2+{\beta_t}^2}{2|\alpha_t-\alpha_r|} .
\end{eqnarray}
On the other hand, $\delta_s$ reflects the deformation of the shadow and is defined as
\begin{eqnarray}
\delta_s=\frac{d_s}{R_s}=\frac{|\alpha_l-\tilde{\alpha_l}|}{R_s} ,
\end{eqnarray}
where $d_s$ represents the distances between the most left point of the black hole shadow and the reference circle.
By definition, these quantities are closely connected to astronomical observations.

\begin{figure}[tbh]
	\centering
	\begin{minipage}[c]{1\linewidth}
		\centering
		\includegraphics[width=5cm]{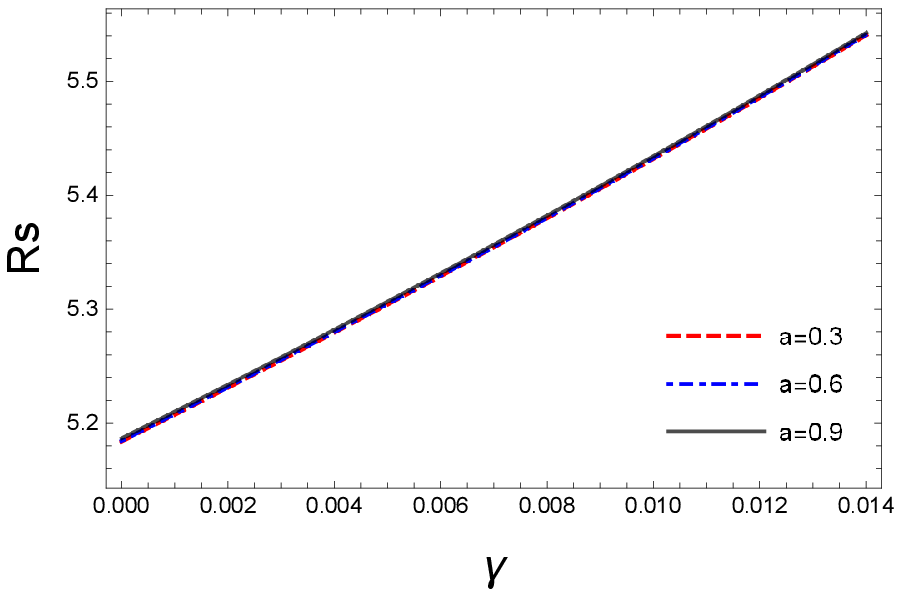}
		\includegraphics[width=5cm]{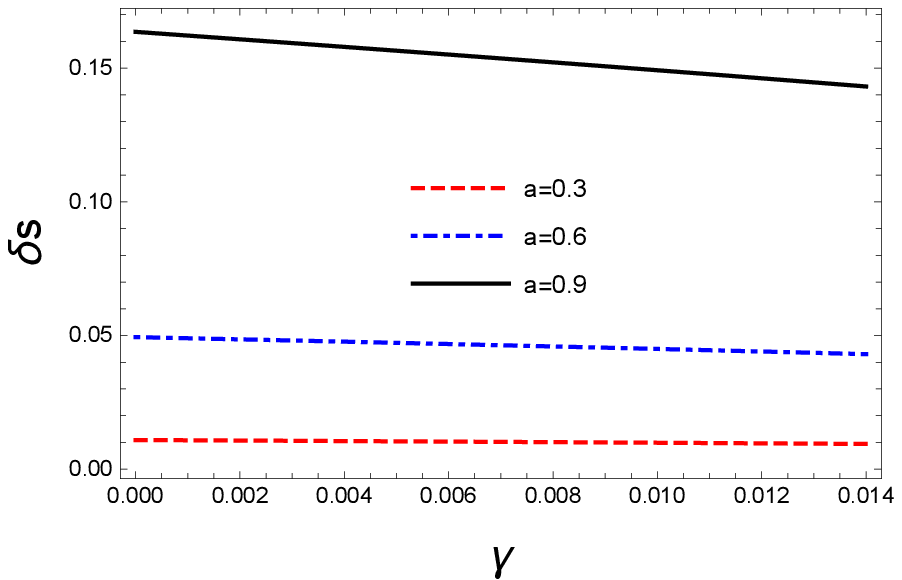}
	\end{minipage}%
	\caption{The obtained shadow radius $R_s$ and distortion parameter $\delta_s$ as functions of the quintessential parameter $\gamma$.
	The calculations are carried out for different values of spin $a$, for given electric charge $Q=0.4$.}
	\label{model2-ob-1}
\end{figure}

\begin{figure}[tbh]
	\centering
	\begin{minipage}[c]{1\linewidth}
		\centering
		\includegraphics[width=5cm]{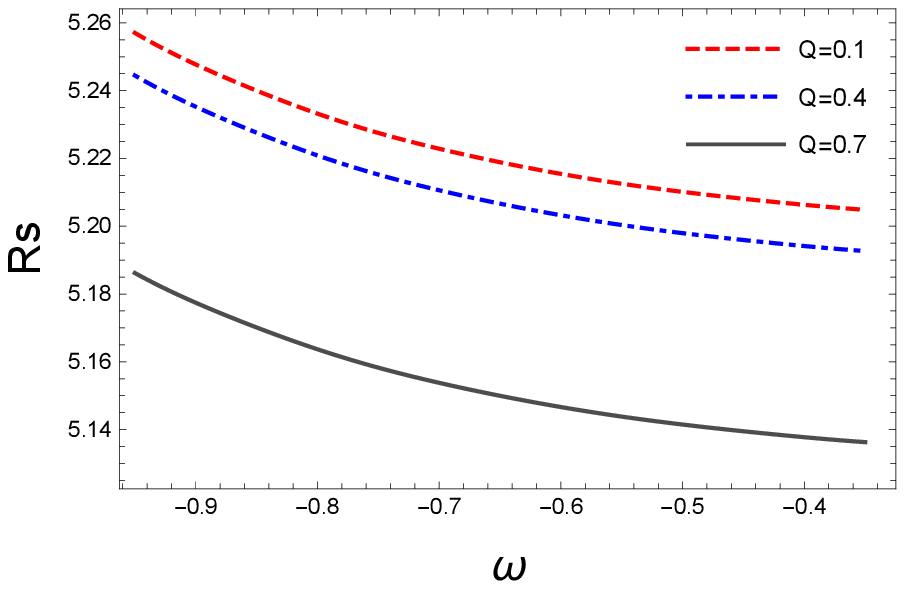}
		\includegraphics[width=5cm]{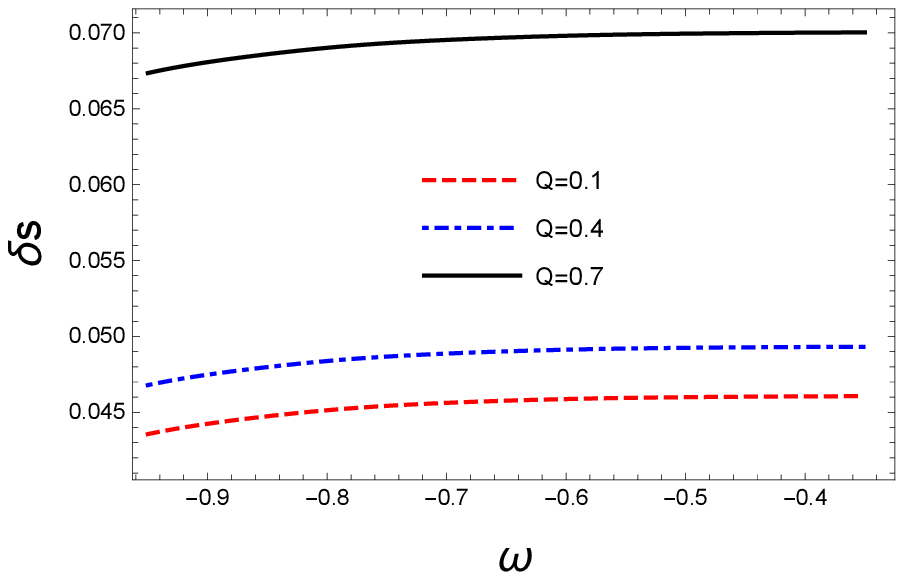}
	\end{minipage}%
	\caption{The obtained shadow radius $R_s$ and distortion parameter $\delta_s$ as functions of state parameter $\omega$.
	The calculations are carried out for different values of charge $Q$, for given spin $a=0.6$ and $\gamma=0.001$.}
	\label{model2-ob-2}
\end{figure}

\begin{figure}[tbh]
	\centering
	\begin{minipage}[c]{1\linewidth}
		\centering
		\includegraphics[width=5cm]{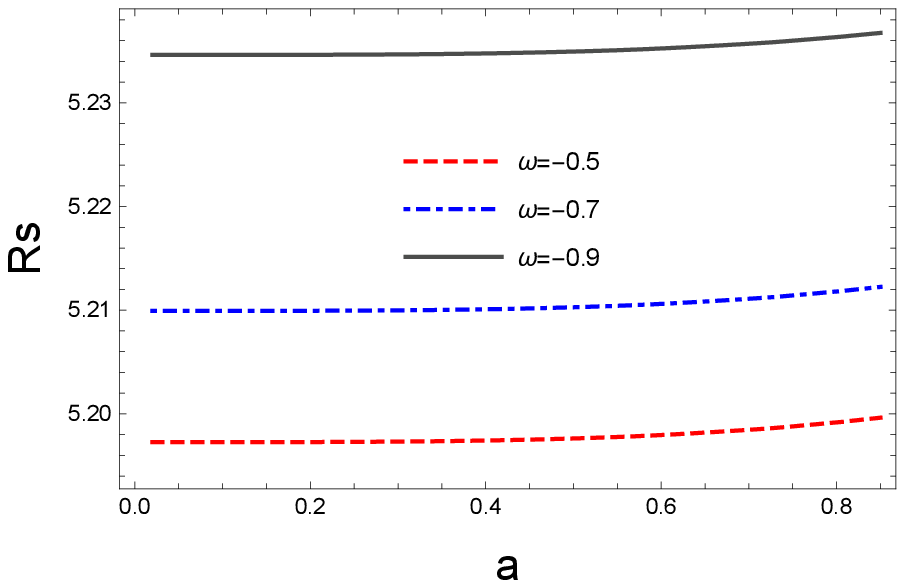}
		\includegraphics[width=5cm]{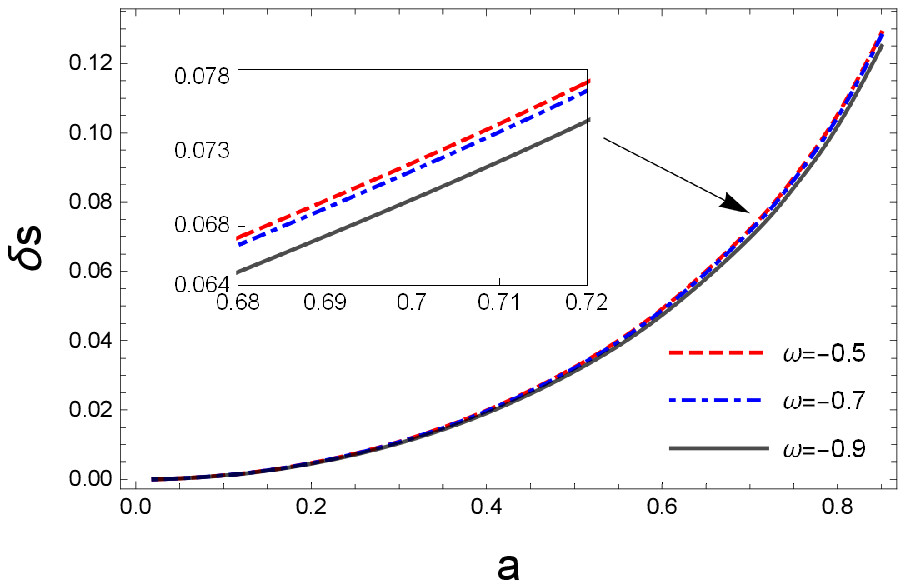}
	\end{minipage}%
	\caption{The obtained shadow radius $R_s$ and distortion parameter $\delta_s$ as functions of state parameter $\omega$.
	The calculations are carried out for different values of state parameter $\omega$, for given magnetic charge $Q=0.4$ and $\gamma=0.001$.}
	\label{model2-ob-34}
\end{figure}

\begin{figure}[htb]
	\centering
	\begin{minipage}[c]{1\linewidth}
		\centering
		\includegraphics[width=5cm]{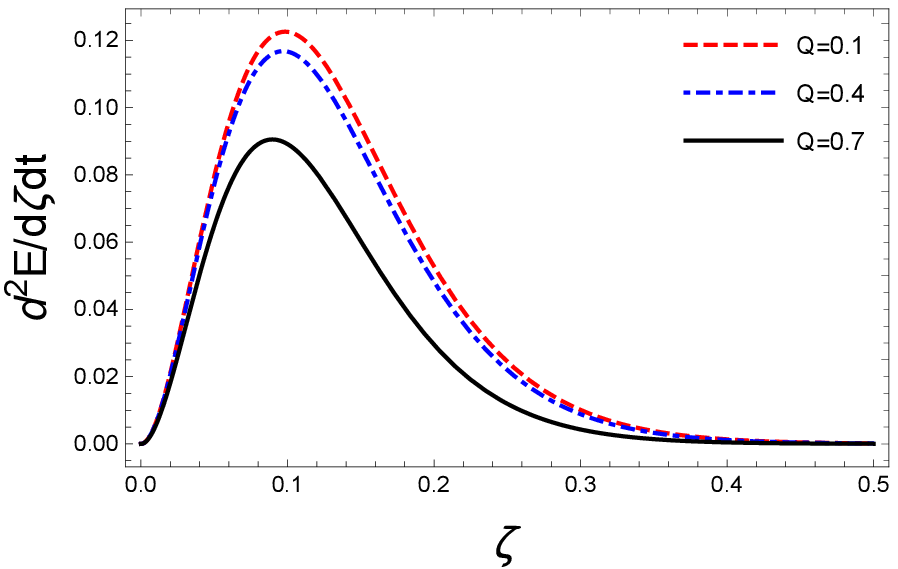}
		\includegraphics[width=5cm]{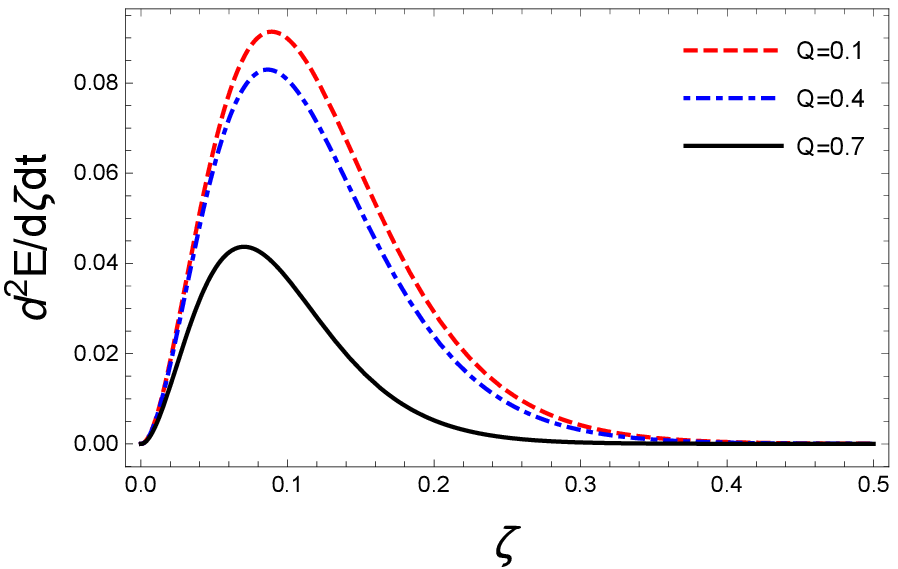}
		\includegraphics[width=5cm]{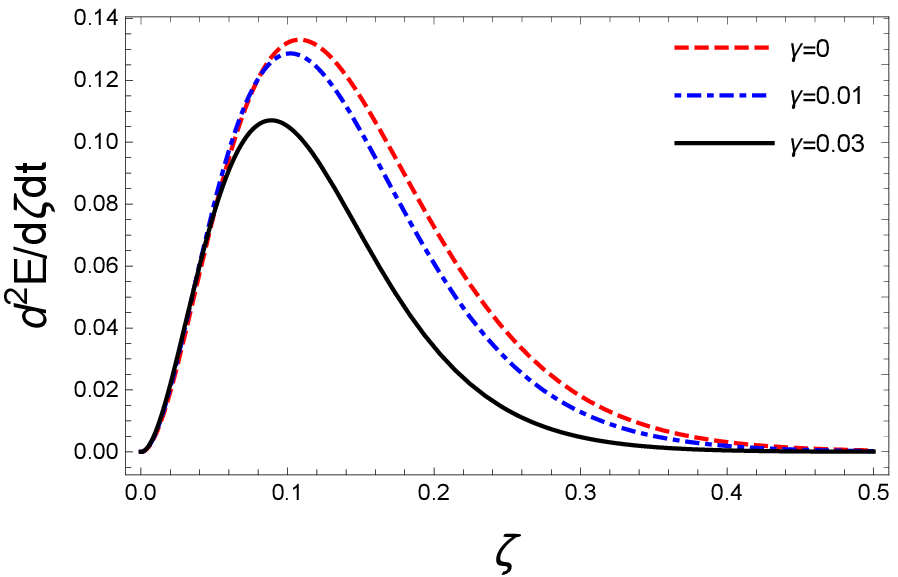}
		\includegraphics[width=5cm]{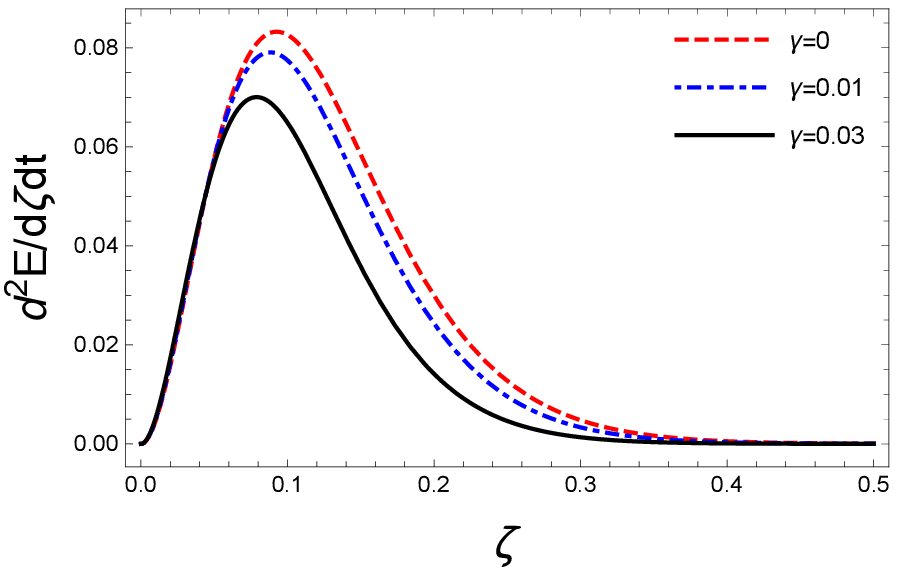}
	\end{minipage}%
	\caption{The obtained evolution of the energy emission rate as a function of the photon frequency $\omega$.
	The calculations are carried out with given $\gamma=0.02$ for the two plots in the upper low and $Q=0.5$ for the two plots in the lower row.
	The plots in the left panel are evaluated using $a=0.1$, and those in the right panel are using $a=0.6$.}
	\label{model2-energy}
\end{figure}

In Fig.~\ref{model2-ob-1}, we show the calculated shadow radius $R_s$ and distortion parameter $\delta_s$ as functions of the quintessential parameter $\gamma$.
The left plot indicates that it is rather difficult to distinguish the black hole spin $a$ by using the observed shadow radius $R_s$ since the curves with different values of $a$ mostly overlap with each other.
However, the shadow radius $R_s$ is shown to be sensitive to the quintessential parameter $\gamma$, as it increases with increasing $\gamma$.
On the other hand, the right plot indicates that the distortion $\delta_s$ is sensitive to $a$, however, it varies slowly as a function of $\gamma$.

To show the effect of the equation of state of the quintessence, in Fig.~\ref{model2-ob-2} we analysize the resulting dependence of $R_s$ and $\delta_s$ on the state parameter $\omega$. 
It is observed that the shadow radius $R_s$ decreases more significantly as the $\omega$ increases, while the distortion $\delta_s$ increases slowly with $\omega$.
By choosing $\omega = -0.5,~-0.7$ and $-0.9$, in Fig.~\ref{model2-ob-34}, we also show the resulting dependence of $R_s$ and $\delta_s$ on the black hole spin. 
One observes that for different values of the spin $a$, the shadow radius $R_s$ is sensitive to $\omega$.
While for the distortion $\delta_s$, its dependence on $\omega$ is hardly observable.

The results obtained in the above figures indicate that the information on the quintessential parameter might be extracted if one can precisely measure the mass and size of the black hole shadow irrelevant to the specific value of the spin.
For specific measurements, the latter, on the other hand, might be sensitively dependent on the deformation of the shadow.
The modifications to the shadows due to the state parameter $\omega$ is found to be of the same order of magnitude to those associated to the spin and charge.
The latter are also shown to be rather sensitive to both the size and distortion of the black hole shadow, potentially related to its significant effect on the horizon structure and photon region, discussed in the previous section.
In particular, if one chooses the observables properly, information on the relevant black hole parameter is expected to be extracted.
Therefore, in the present model, the state parameter does seem to play a significant role in determining the main features of the shadow.

From the perspective of an observer at infinity, the shadow of a black hole can be viewed as a high-energy absorption cross-section.
In the case of a spherically symmetric black hole, the absorption cross-section oscillates around a limiting constant value $\sigma_{lim}$, which is the same as the geometrical cross-section of photon sphere~\cite{Misner:1973prb}.
Therefore, this value can be expressed as~\cite{Wei:2013kza}
\begin{eqnarray}
\sigma_{lim}\approx\pi{R_s^2},
\end{eqnarray}
where $R_s$ is the shadow radius of a black hole. 
By using this limiting constant value, the energy emissiom rate can be shown to possess the following form
\begin{eqnarray}
\frac{d^2E(\zeta)}{d{\zeta}dt}=\frac{2\pi^2\sigma_{lim}}{e^{{\zeta}/{T}}-1}{\zeta}^3 ,
\end{eqnarray}
where $\zeta$ represents the frequency of photon, and $T$ is the Hawking temperature of a black hole for the outer event horizon, which reads
\begin{eqnarray}
T=\lim_{\theta\rightarrow0,r\rightarrow{r_+}}\frac{\partial_r \sqrt{-g_{tt}}}{2\pi\sqrt{g_{rr}}},
\end{eqnarray}
where $r_+$ denotes the outer event horizon of the black hole. 
Thus, the Hawking temperature of a rotating black hole immersed in the quintessence field is found to be
\begin{eqnarray}
T_{\mathrm{Quintessence}}=\frac{r_+^2 f'(r_+)(r_+^2+a^2)+2a^2r_+(f(r_+)-1)}{4\pi(r_+^2+a^2)^2} ,
\end{eqnarray}
where $f(r)=1-\frac{2Mr^2}{r^3+Q^3}-\gamma r$.

The evolution of energy emission rate as a function of the photon frequency $\zeta$ for the present metric is calculated and shown in Fig.~\ref{model2-energy}.
One observes that the peak decreases and shifts to lower frequency with increasing $a$, $Q$, and $\gamma$.
The result is largely consistent with the finding for Kerr black holes.

\section{Conclusions}\label{summary}

To summarize, in this work, we studied the optical properties of a class of magnetically charged rotating black hole spacetimes.
The black holes are surrounded by the quintessence field, and subsequently, the resulting black hole shadows are found to be modified by the presence of dark energy.
We investigated the photon region and the black hole shadow, as well as their dependence on the relevant physical conditions.
In particular, the effects of the state parameter of the quintessence, the angular momentum, and the magnitude of the magnetic charge are explored.
It is shown that the photon regions sensitively depend on the horizon structure and possess intricate features.
Moreover, from the viewpoint of a static observer, we explore a few physical observables which are associated with the distortion of the observed black hole shadows.
It is found that the presence and the properties of the dark energy might be implied from the empirical studies of the black hole shadows, as a few relevant physical quantities substantially affect the size of the shadow in a fashion independent of the rotation of the black hole.

\section*{Acknowledgments}

This research is supported by by National Key R\&D Program of China under Grant No. 2020YFC2201400, and the Major Program of the National Natural Science Foundation of China under Grant No. 11690021 and the National Natural Science Foundation of China under Grant No. 11505066.
We also gratefully acknowledge the financial support from
Funda\c{c}\~ao de Amparo \`a Pesquisa do Estado de S\~ao Paulo (FAPESP), 
Funda\c{c}\~ao de Amparo \`a Pesquisa do Estado do Rio de Janeiro (FAPERJ), 
Conselho Nacional de Desenvolvimento Cient\'{\i}fico e Tecnol\'ogico (CNPq), 
Coordena\c{c}\~ao de Aperfei\c{c}oamento de Pessoal de N\'ivel Superior (CAPES).

\end{document}